\documentclass[journal=pasa, manuscript=research-paper, year=2024, volume=37]{cup-journal}

\usepackage{graphicx}
\usepackage{epsfig}
\usepackage{amsmath}
\usepackage{amssymb}
\usepackage{rotating}
\usepackage{color}
\usepackage{pifont}
\usepackage{longtable}
\usepackage{pdflscape,lscape}
\usepackage{appendix}
\usepackage[normalem]{ulem}
\usepackage[utf8x] {inputenc}
\usepackage[bookmarks=false]{hyperref}
\newcommand{\teff}{$T_{\rm eff}$}
\newcommand{\logg}{$\log g$}
\newcommand{\vsini}{$v \sin i$}
\newcommand{\kms}{km\,s$^{-1}$}
\newcommand{\ds}{$\delta$\,Scuti}
\newcommand{\bc}{$\beta$\,Cephei}
\newcommand{\gd}{$\gamma$\,Doradus}

\newcommand\simgt{\ {\raise-.5ex\hbox{$\buildrel>\over\sim$}}\ }
\newcommand\simlt{\ {\raise-.5ex\hbox{$\buildrel<\over\sim$}}\ }
\usepackage{microtype,,siunitx,booktabs}
\title{On the Existence of ``Maia variables''}

\author{F. Kahraman Ali\c{c}avu\c{s}}
\affiliation{\c{C}anakkale Onsekiz Mart University, Science Faculty, Physics Department, 17100, Canakkale, T\"{u}rkiye}
\email[F. Kahraman Ali\c{c}avu\c{s} \& G. Handler]{filizkahraman01@gmail.com, gerald@camk.edu.pl }
\alsoaffiliation{\c{C}anakkale Onsekiz Mart University, Astrophysics Research Center and Ulup{\i}nar Observatory, TR-17100, Çanakkale, T\"{u}rkiye}

\author{G. Handler}
\affiliation{Nicolaus Copernicus Astronomical Center, Polish Academy of Sciences, Bartycka 18, PL-00-716 Warsaw, Poland}

\author{S. Chowdhury}
\affiliation{Nicolaus Copernicus Astronomical Center, Polish Academy of Sciences, Bartycka 18, PL-00-716 Warsaw, Poland}

\author{E. Niemczura}
\affiliation{University of Wroc\l{}aw, Astronomical Institute, Kopernika 11, 51-622, Wroc\l{}aw, Poland}

\author{R. Jayaraman}
\affiliation{Department of Physics, and Kavli Institute for Astrophysics and Space Reserch, M.I.T., Cambridge, MA 02139, USA}

\author{P. De Cat}
\affiliation{Royal Observatory of Belgium, Ringlaan 3, B-1180 Brussel, Belgium}

\author{D. Ozuyar}
\affiliation{Ankara University, Faculty of Science, Dept. of Astronomy and Space Sciences, 06100, Tandogan, Ankara, T\"{u}rkiye}

\author{F. Ali\c{c}avu\c{s}}
\affiliation{\c{C}anakkale Onsekiz Mart University, Science Faculty, Physics Department, 17100, Canakkale, T\"{u}rkiye}

\alsoaffiliation{\c{C}anakkale Onsekiz Mart University, Astrophysics Research Center and Ulup{\i}nar Observatory, TR-17100, Çanakkale, T\"{u}rkiye}

\doi{}

\received {dd Mmm YYYY}
\revised  {dd Mmm YYYY}
\accepted {dd Mmm YYYY}
\published{dd Mmm YYYY}


\begin{document}

\begin{abstract}
There are different classes of pulsating stars in the H-R diagram. While many of those classes are undisputed, some remain a mystery such as the objects historically called "Maia variables". Whereas the presence of such a class was suggested seven decades ago, no pulsational driving mechanism is known that could excite short-period oscillations in these late B to early A-type stars. Alternative hypotheses that would render the reports of variability of those stars erroneous have been proposed such as incorrect effective temperatures, binarity or rapid rotation, but no certain conclusions have been reached yet. Therefore the existence of these variables as a homogeneous class of pulsating star is still under discussion. Meanwhile, many new candidates of these variables have been claimed especially by using photometric observations of space telescopes. In this study, we examined 31 objects that are alleged members of this hypothetical group and carried out detailed spectroscopic and photometric analyses to test the proposed hypotheses for their cause of variability. The \teff, \logg, \vsini, and chemical abundances of the targets were determined and the TESS photometric data were examined. As a result, we found that most of these targets are located inside the \ds, \bc, or SPB star instability strips, a few show evidence for binarity and others for rapid rotation. We give arguments that none of the apparently rapid pulsations in our targets is caused by a star outside any known instability strip. By extrapolation, we argue that most stars proposed as pulsators outside well-established instability domains are misclassified. Hence there is no sufficient evidence justifying the existence of a class of pulsating stars formerly known as the "Maia variables".
\end{abstract}

\section{INTRODUCTION}\label{sec:int}
Perhaps the most efficient and accurate way to investigate stellar interior structure is asteroseismology which uses the eigenfrequencies of stellar oscillations to probe the interiors of stars. Using asteroseismology phenomena such as internal rotation, angular momentum, and convective core overshooting have been successfully examined \citep[e.g.,][]{2018A&A...618A..47C, 2018MNRAS.475..879T, 2019MNRAS.485.3544W, 2022ARA&A..60...31K}. 

The loci of several different classes of pulsating stars can be delineated in the Hertzsprung-Russell (H-R) diagram. These loci are called "instability strips" and comprise groups of different evolutionary statuses. Among those on the main sequence, six types of pulsating stars are commonly known. Moving from hotter to cooler temperatures, these variables are the $\beta$\,Cephei stars, the slowly pulsating B-type (SPB) stars, the $\delta$\,Scuti stars, the rapidly oscillating Ap (roAp) stars, the $\gamma$\,Doradus stars and solar-like oscillators. The \bc\, stars are objects of spectral types B0 -- B3 showing low-order pressure (p) and gravity (g) modes with frequencies $\simgt$3.5\,d$^{-1}$ \citep{2005ApJS..158..193S}. On the other hand, the SPB variables are B3 -- B9 stars which exhibit high-order g-mode oscillations with frequencies $\simlt$3.5\,d$^{-1}$ \citep{2010aste.book.....A}. The \ds\, stars exhibit high-frequencies ($\simgt$5\,d$^{-1}$) similar to \bc\, variables, however, their spectral types range from A2 to F5 \citep{2013AJ....145..132C, 2011A&A...534A.125U}. The roAp stars comprise similar spectral types, but are chemically peculiar due to the presence of strong magnetic fields and pulsate in high-order pressure modes with even higher frequencies ($\simgt$50\,d$^{-1}$) \citep{2024MNRAS.527.9548H}. The \gd\, stars are low-frequency ($\simlt$5\,d$^{-1}$) g mode pulsators with spectral types of F2 -- F6 \citep{2011A&A...534A.125U}. Solar-like oscillators are of even later spectral type on the main sequence and show stochastic high-order p mode oscillations with frequencies around 300 d$^{-1}$ \citep[e.g.,][]{2021FrASS...7..102J}. All these pulsators have their own instability strips where theoretical models can explain the observed types of oscillations. Consequently, almost the whole main sequence is populated with pulsating stars with spectral types ranging from early B to late G, but with a marked gap between spectral types B9 and A2. The reason for this gap is that both B-type variable star classes show pulsations triggered by the $\kappa$ mechanism operating in the partial ionization zone of iron group elements \citep{1993MNRAS.265..588D}, but the $\delta$ Scuti stars are triggered by the $\kappa$ and/or the turbulent mechanism acting in the partial ionization zones of Helium \citep{1962ZA.....54..114B, 2014ApJ...796..118A} and hydrogen \citep[e.g.,][]{2013MNRAS.436.1639C}. In the gap between those groups of pulsators at late B/early A spectral type, none of the currently known pulsation mechanisms are able to drive pulsations in models, consistent with observations. In particular, no pulsations with intrinsic frequencies in excess of 3 d$^{-1}$ should exist in stars with spectral types between B3 and A2. 

However, a group of high-frequency pulsating stars in this gap and in the cool part of the SPB instability strip has been proposed by several authors \citep[e.g.,][]{1955S&T....14..461S, 1995A&A...300..783L}.
These variables seem to have high frequencies like \bc\, and \ds\, stars but they are too cool to be a \bc\, and too hot to be a \ds\, pulsator. Hence they appear to populate the gap where pressure mode pulsations should not exist. These variables have historically been known as the "Maia stars" following the apparent prototype of these variables, Maia in the Pleiades cluster. Maia was first thought to be a high-frequency pulsator \citep{1955S&T....14..461S}, but this claim was refuted several times \citep[e.g.,][and references therein]{1985ApJ...289..213M} and in the end it was shown that Maia is actually a rotational variable \citep[HgMn star, ][]{2017MNRAS.471.2882W} without any detectable pulsational variability. However, recent studies, especially those based on space-based data, revived interest in those objects \citep{2016MNRAS.460.1318B, 2020MNRAS.493.5871B,2023FrASS..1066750B}. 

Furthermore, examples of objects with slower apparently pulsational variations have been reported in the gap between the main sequence instability strips \citep[e.g.,][]{2005A&A...431..615A,2009A&A...506..471D,2013A&A...554A.108M, 2022MNRAS.515..828S}, which are not to be confused with the short period variables mentioned above. Nevertheless, the hypotheses to explain all these variable stars are essentially the same. We call the first one the "rapid rotation hypothesis". This idea would explain the apparent pulsators as rapidly rotating SPB or \bc\, stars seen equator-on. Because of this fast rotation, the star would have an ellipsoidal-like shape, and its effective temperature (\teff) would be lower near the equator and higher near the poles. This would have several consequences. First, the observationally determined \teff\, would be different from the overall value due to projection effects, modifying the star's position in the HR Diagram. Second, the rapid rotation would shift the observed pulsational mode frequencies \citep{2005MNRAS.360..465T, 2014A&A...569A..18S,2017MNRAS.469...13S, 2016A&A...595L...1M} to higher values. Third, Rossby waves (r modes) could be excited \citep{2017MNRAS.469...13S,2018MNRAS.474.2774S}. Fourth, the non-uniform temperature distribution in the star could facilitate pulsational excitation near the stellar poles or the equator \cite[cf.][]{2023MNRAS.521.4765K}. Fifth, rapid rotation would extend the SPB instability strip, also towards lower temperatures \citep{2005MNRAS.360..465T}. Another hypothesis to explain pulsation outside of any known instability strip is associated with binarity (hereinafter the "binary hypothesis"). The pulsating stars could be components of binary systems consisting of a non-pulsating late B-type star, which dominates the spectrum, and a \ds\, pulsator. Given that most B-type stars have companions, and often more than one \citep{2017ApJS..230...15M}, such a scenario is not unlikely. Therefore, apparent p-mode pulsations would be associated with a star too hot to have them excited. The third hypothesis involves simple observational uncertainties that lead to incorrect \teff\, values (hereinafter the "wrong \teff\, hypothesis"). Most of the variables under question were observed by the Transiting Exoplanet Survey Satellite \citep[{\it {\it TESS}},][]{2014SPIE.9143E..20R}. The \teff\, of these B type systems given in the {\it {\it TESS}} input catalogue \citep{2011AJ....142..112B, 2019AJ....158..138S} may not always be reliable. These \teff\, parameters are in most cases photometric estimates based on a system lacking a U-band filter, and that carry some uncertainties regarding dereddening for stars located close to the Galactic plane \citep{2019AJ....158..138S}. Therefore the \teff\, values for hot stars from the TIC catalogue have to be treated with caution. Finally, regarding {\it TESS} data, a "contamination hypothesis" must be considered. Due to the large (21") size of {\it TESS}'s CCD pixels on the sky, the light of nearby stars may fall into the photometric aperture, hence, for the purpose of example, an unrelated \ds\, star may introduce short-period pulsational variations into the light curve of a star outside of any pulsational instability strip.

\cite{2016MNRAS.460.1318B, 2020MNRAS.493.5871B} tested some of these hypotheses, but without an adequate amount of good quality spectroscopic data. In a recent study \cite{2023FrASS..1066750B} even suggested that Maia variables are an extension of \ds\, stars and there is an interplay between available pulsational driving mechanisms, but this conclusion was partly based on inhomogeneous literature data. To improve on this situation one needs to study these variable stars with medium to high-resolution spectroscopic data to determine some important parameters (e.g. \teff, $v \sin i$) accurately. 

In a recent study, \cite{2023MNRAS.521.4765K} analyzed an apparently genuine, very rapidly rotating p-mode pulsator of spectral type A0 (HD\,42477), which is located between the \bc\, and \ds\, instability strips, by taking into account the incorrect \teff, contamination, and binarity hypotheses. They ruled out or placed strong constraints against all these hypotheses and concluded that p-mode pulsations can be present in the mentioned region of the H-R diagram. Their excitation would be explained by coupling with several g and r modes or by the classical $\kappa-\gamma$ mechanism operating in the He{\sc ii} ionization zone in the stellar equatorial regions, hence is understood based on current astrophysical knowledge.

Recapitulating, there is one well-documented case of pressure mode pulsation between the instability strips of the SPB and \ds\, stars. In addition, there are claims of hundreds of such stars based on space photometry data, but without proper assessments of whether the fast rotation, the binary, the contamination, or the wrong \teff\, hypotheses would apply. Given the sheer number of these objects, is it really possible that most of them were claimed by misfortune?

In this study, we present detailed analyses of some candidate variables in the gap between the main sequence instability strips to test the proposed hypotheses and to establish whether a class of "Maia variables"\footnote{Given that Maia itself is not a pulsating star, a different name for this prospective class would be required.} is justified. We use {\it {\it TESS}} photometric data and medium and/or high-resolution spectra and the paper is organized as follows. In Sect.\,2, we introduce the target selection and observations. In Sect.\,3, the spectroscopic analyses including radial velocity measurements, determination of atmospheric parameters, and chemical abundances are presented. A contamination analysis on {\it {\it TESS}} data is introduced in Sect.\,4. Discussions and conclusions are given in Sect.\,5, and 6, respectively. 

\section{Target selection and observations}
Our variable star candidates were selected from the list of \cite{2020MNRAS.493.5871B}. These authors defined priority classes considering the methods used to estimate \teff\, values. The highest priority was given for spectroscopic \teff\, while the lowest priority was defined for \teff\, estimated from the spectral types. In between, there are \teff\, estimations from some photometric colours. In this study, we preferentially chose stars of lower priority, as this adds more value to our spectral analyses for accurate \teff\, determinations. 

\begin{figure}
\includegraphics[width=8.58cm, angle=0]{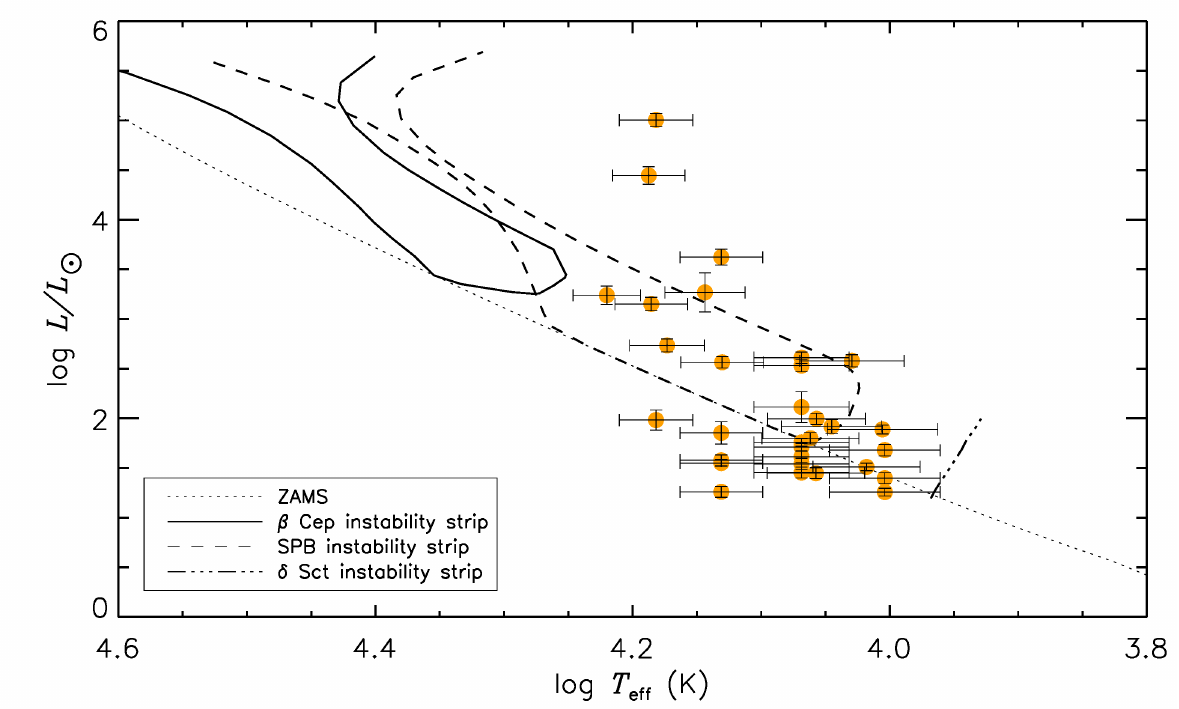}
\caption{The initial positions of the selected candidates in the H-R diagram as derived by \cite{2020MNRAS.493.5871B}. The theoretical instability strips of the \bc\, and SPB stars were taken from \citet{1999AcA....49..119P}. The domain of the \ds\, stars was taken from \cite{2019MNRAS.485.2380M}.}
\label{figure1_first_position}
\end{figure}

We reanalyzed the {\it {\it TESS}} Simple Aperture Photometry (SAP) 2-min cadence light curves of our selected candidates to confirm the presence of pressure mode oscillations in these variables. We made use of the Period04 \citep{2005CoAst.146...53L} package, which utilizes simultaneous multi-frequency sine-wave fitting with input parameter optimization. After removing the sine-wave fits from the original data, the residuals were then re-analyzed in the same manner to detect further periodicities. The signals were extracted until we reached a Signal-to-Noise ratio (SNR) of 4.5 \citep{2021AcA....71..113B}. In addition, the initial positions of the selected candidates in the H-R diagram were checked using the luminosity ($L$) values calculated from the Gaia parallaxes \citep{2022yCat.1355....0G}, interstellar reddening, E(B-V), calculated from the interstellar extinction map \citep{2005AJ....130..659A}, the bolometric corrections \citep{1996ApJ...469..355F} and the \teff\, values given by \cite{2020MNRAS.493.5871B} (see, Fig.\,\ref{figure1_first_position}). As a result, 31 candidates were selected for this study; they are listed in Table\,\ref{table1_info}.

\begin{table}
    \centering
\caption{The list of the targets. The number of the taken spectra from FIES and HERMES and their signal-to-noise ratios (SNR) are given in the last two columns, respectively. These spectra were taken between 2020 and 2021. 
} \label{table1_info}
    \begin{tabular}{llccl}
    \hline
    TIC & HD & V                  &\#FIES/HERMES  &  SNR   \\ 
        &    & (mag) & Spectra     &  FIES/HERMES      \\
\hline
16781787& 207674 &8.26& 1\,/\,1   & 92 \,/\, 73 \\ 
20534584& 176052 &8.41& $-$\,/\,2 & $-$ \,/\,  74, 78  \\ 
26368855& 222017 &8.93& 2\,/\,2   & 89, 38 \,/\, 70, 73  \\ 
26369887& 222142 &9.55& 1\,/\,1   & 49 \,/\, 53 \\ 
27804376& 210661 &7.53& 2\,/\,1   & 136, 65 \,/\, 103  \\ 
36557487& 49643  &5.75& $-$\,/\,2 & $-$ \,/\, 123, 126\\ 
50850670&        &9.09& 2\,/\,2   & 73, 49 \,/\, 30, 27  \\ 
64774351& 204905 &7.24& 3\,/\,2& 138, 141, 140 \,/\, 96, 16  \\
83803744&  1009  &8.38& 2\,/\,$-$ & 71, 30\,/\,  $-$\\ 
90647390& 278751 &10.36& 2\,/\,2   & 36, 29 \,/\, 13, 38  \\ 
192780911& 19921 &7.73& 1\,/\,2   & 145 \,/\, 56, 99   \\ 
201739321&  356  &8.01& 3\,/\,1   & 88, 24, 58 \,/\, 96  \\ 
202567458& 332044&9.87& 2\,/\,$-$ & 67, 70 \,/\, $-$ \\
236785664& 175511&6.94& $-$\,/\,4 & $-$ \,/\, 86, 82, 141, 133  \\ 
251196433&       &10.12& 1\,/\,2   & 47 \,/\, 48, 45  \\ 
255996252& 196358&8.34& 3\,/\,2   & 28, 59, 81 \,/\, 87, 86  \\
312637783& 228456&10.02& 3\,/\,$-$ & 51, 56, 59 \,/\, $-$ \\ 
317512446& 220855&8.64& 2\,/\,1   & 83, 74 \,/\, 15  \\ 
335484421&       &8.73& 2\,/\,$-$ & 100, 78 \,/\, $-$  \\ 
352781227&351435 &11.24& $-$\,/\,2 & $-$ \,/\, 23, 23  \\
354793407& 21699 &5.46& $-$\,/\,2 & $-$ \,/\, 136, 112  \\ 
367351021& 216486&8.97& 2\,/\,$-$ & 74, 80 \,/\, $-$ \\ 
374038418&       &9.61& 2\,/\,2   & 38, 34 \,/\, 48, 38  \\ 
377099704& 177195&8.74& $-$\,/\,5 & $-$ \,/\, 60, 66,  73, 70,  61 \\
377443211& 185757&8.31& 2\,/\,1   & 62, 84 \,/\, 87 \\ 
387757610&200506 &8.13& 2\,/\,2   & 89, 85 \,/\, 97, 20  \\ 
402910080& 193989&8.57& 2\,/\,$-$ & 116, 108 \,/\, $-$ \\
424032547& 350990&10.32& 2\,/\,$-$ & 52, 50 \,/\, $-$\\ 
434340780&       &9.14& 2\,/\,$-$ & 89, 90 \,/\, $-$\\ 
450085823&287149 &10.14& 1\,/\,$-$ & 35 \,/\, $-$\\
467219052&239419 &9.77& 2\,/\,1 & 52, 40 \,/\, 63\\
    \hline
    \end{tabular}
\end{table}

For these candidates, we carried out spectroscopic surveys. Some spectra were taken from the Fibre-fed \'{E}chelle Spectrograph \citep[FIES, ][]{2014AN....335...41T} that is a cross-dispersed high-resolution \'{e}chelle spectrograph attached to the 2.56-m Nordic Optical Telescope of the Roque de los Muchachos Observatory (ORM, La Palma, Spain). It has three different resolution options; low ($R=25000$), medium ($R=46000$) and high ($R=67000$). Considering the \teff\, values and the V-band magnitudes of the target stars, the low-resolution mode was used during the observations to achieve a higher signal-to-noise ratio (SNR). In addition, this resolving power is good enough for spectral analysis of systems having medium to high projected rotational velocity (\vsini $\gtrsim$30 \kms). We also gathered some spectra from the High Efficiency and Resolution Mercator \'{E}chelle Spectrograph \citep[HERMES, ][]{2011A&A...526A..69R} which is a high-resolution ($R=85000$) fibre-fed \'{e}chelle spectrograph attached to the 1.2-m Mercator telescope at the Roque de los Muchachos Observatory (ORM, La Palma, Spain). Some stars have spectra from both spectrometers, while some have only from one. For each target, we tried to obtain at least two spectra taken at different times to check for possible binary-induced radial velocity shifts. The information about the spectroscopic surveys is given in Table\,\ref{table1_info}. 

\begin{table*}
    \centering
\caption{Atmospheric parameters of the stars determined from the hydrogen Balmer lines and the updated parameters. An asterisk represents a fixed \logg\, value for TIC\,352781227 that we could not determine because of rapid rotation and low SNR. Since some stars are not suitable for further analysis because of their large rotational velocities and low SNR values of their spectra, only some of the parameters could be updated (see, Sect\,\ref{chemical}). $E(B-V)$ was determined from the sodium lines and the final $L$ parameters are given. The Gaia RUWE parameters are also listed in the last column.} \label{table2_atmpars}
\begin{tabular}{lcc|cccr|cc|c}
    \hline
    &\multicolumn{2}{c|}{\hrulefill Balmer lines\,\hrulefill}  &\multicolumn{4}{c}{\hrulefill \,Updated Parameters\,\hrulefill}&&            &\\
    TIC  &  \teff    & \logg   & \teff & \logg & $\xi$ & \vsini   & E(B-V)  &log\,$L$/$L_{\odot}$ & RUWE \\
             &  (K)        & (cgs)              & (K)     & (cgs)   &(\kms) & (\kms)     & (mag\,$\pm$\,0.02)  &                    & \\
\hline
    16781787 & 11000\,$\pm$\,200  & 3.8\,$\pm$\,0.1 &                  &                 &                 &260\,$\pm$\,22 & 0.15 & 2.042\,$\pm$\,0.066 & 1.12\\ 
    20534584 & 9600\,$\pm$\,200   & 3.6\,$\pm$\,0.1 & 9600\,$\pm$\,100 & 3.5\,$\pm$\,0.1 & 2.5\,$\pm$\,0.2&150\,$\pm$\,10 & 0.03 & 1.841\,$\pm$\,0.043 & 2.26\\ 
    26368855 & 8500\,$\pm$\,100   & 4.0\,$\pm$\,0.1 & 8700\,$\pm$\,100 & 4.1\,$\pm$\,0.1 & 2.5\,$\pm$\,0.2 & 98\,$\pm$\,5  & 0.07 & 1.366\,$\pm$\,0.055 & 1.29\\ 
    26369887 & 10000\,$\pm$\,500  & 3.5\,$\pm$\,0.1 &                  &                 &                 & 230\,$\pm$\,10& 0.05 & 1.469\,$\pm$\,0.055 & 0.99\\
    27804376 & 8500\,$\pm$\,100   & 4.0\,$\pm$\,0.1 & 8800\,$\pm$\,100 & 4.0\,$\pm$\,0.1 & 1.5\,$\pm$\,0.2 & 135\,$\pm$\,10& 0.07 & 1.288\,$\pm$\,0.054 & 1.15\\
    36557487 & 14000\,$\pm$\,500  & 4.0\,$\pm$\,0.1 &                  &                 &                 & 260\,$\pm$\,17& 0.04 & 3.273\,$\pm$\,0.200 & 11.59\\ 
    50850670 & 7400\,$\pm$\,100   & 4.0\,$\pm$\,0.1 & 7600\,$\pm$\,100 & 4.0\,$\pm$\,0.1 & 2.9\,$\pm$\,0.2 & 79\,$\pm$\,3  & 0.03 & 0.818\,$\pm$\,0.053 & 1.08\\ 
    64774351 & 10000\,$\pm$\,200  & 4.0\,$\pm$\,0.1 &                  &                 &                 & 230\,$\pm$\,14& 0.01 & 1.867\,$\pm$\,0.050 & 0.89\\ 
    83803744 & 12000\,$\pm$\,300  & 3.6\,$\pm$\,0.1 & 9800\,$\pm$\,200 & 4.0\,$\pm$\,0.1 & 2.5\,$\pm$\,0.2 & 21\,$\pm$\,4  & 0.25 & 2.443\,$\pm$\,0.057 & 1.13\\ 
    90647390 & 7800\,$\pm$\,100   & 4.0\,$\pm$\,0.1 & 8100\,$\pm$\,200 & 4.1\,$\pm$\,0.1 &2.8\,$\pm$\,0.2  & 42\,$\pm$\,4  & 0.18 & 1.384\,$\pm$\,0.101 & 4.25\\ 
    192780911 & 12000\,$\pm$\,500 & 3.5\,$\pm$\,0.1 &                  &                 &                 &273\,$\pm$\,13 & 0.20 & 2.612\,$\pm$\,0.060 & 1.77\\ 
    201739321 & 9000\,$\pm$\,100  & 4.0\,$\pm$\,0.1 & 9000\,$\pm$\,100 & 4.1\,$\pm$\,0.1 &2.6\,$\pm$\,0.2  &119\,$\pm$\,4  & 0.04 & 1.609\,$\pm$\,0.054 & 1.12\\ 
    202567458 & 21000\,$\pm$\,1000& 3.6\,$\pm$\,0.1 &                  &                 &                 &250\,$\pm$\,22 & 0.64 &  3.983\,$\pm$\,0.062& 1.23\\
    236785664 & 12500\,$\pm$\,500 & 4.0\,$\pm$\,0.1 &                  &                 &                 &260\,$\pm$\,20 & 0.02 & 2.579\,$\pm$\,0.058 & 1.20\\ 
    251196433 & 19000\,$\pm$\,1000& 3.5\,$\pm$\,0.1 &                  &                 &                 &255\,$\pm$\,15 & 0.15 & 2.590\,$\pm$\,0.087 & 2.01\\ 
    255996252 & 8500\,$\pm$\,200  & 3.6\,$\pm$\,0.1 & 8600\,$\pm$\,100 & 3.9\,$\pm$\,0.1 & 3.6\,$\pm$\,0.2 &98\,$\pm$\,5   & 0.02 & 1.233\,$\pm$\,0.055 & 0.91\\
    312637783 & 21000\,$\pm$\,1000& 3.8\,$\pm$\,0.2 &22000\,$\pm$\,1000& 3.5\,$\pm$\,0.2 & 1.5\,$\pm$\,0.2 &19\,$\pm$\,2   & 0.40 & 3.725\,$\pm$\,0.091 & 3.53\\ 
    317512446 & 9000\,$\pm$\,100  & 4.0\,$\pm$\,0.1 & 8900\,$\pm$\,100 & 4.1\,$\pm$\,0.1 & 3.0\,$\pm$\,0.2 &40\,$\pm$\,2   & 0.05 & 1.363\,$\pm$\,0.053 & 0.94\\ 
    335484421 & 7000\,$\pm$\,100  & 4.0\,$\pm$\,0.1 & 7300\,$\pm$\,100 & 4.0\,$\pm$\,0.1 & 3.2\,$\pm$\,0.2&56\,$\pm$\,3   & 0.03 & 1.170\,$\pm$\,0.054 & 0.93\\ 
    352781227 & 9500\,$\pm$\,500  & 4.0*            &                  &                 &                 & $\gtrsim$270  & 0.13 & 1.890\,$\pm$\,0.160 & 0.91\\
    354793407 & 14000\,$\pm$\,1000 & 3.6\,$\pm$\,0.1&16000\,$\pm$\,1000& 3.5\,$\pm$\,0.2 & 2.2\,$\pm$\,0.2 & 38\,$\pm$\,4  & 0.04 & 2.825\,$\pm$\,0.061 & 1.81\\ 
    367351021 & 7700\,$\pm$\,100  & 4.0\,$\pm$\,0.1 & 7700\,$\pm$\,100 & 4.0\,$\pm$\,0.1 & 2.7\,$\pm$\,0.1 &77\,$\pm$\,5   & 0.09 & 1.583\,$\pm$\,0.095 & 20.59\\ 
    374038418 & 21000\,$\pm$\,1000 & 3.4\,$\pm$\,0.1&20000\,$\pm$\,1000& 3.5\,$\pm$\,0.1 &2.5\,$\pm$\,0.2  & 55\,$\pm$\,5  & 0.73 & 4.508\,$\pm$\,0.063 & 0.97\\ 
    377099704 & 19000\,$\pm$\,1000& 3.6\,$\pm$\,0.1 &19000\,$\pm$\,1000& 3.5\,$\pm$\,0.2 &2.5\,$\pm$\,0.2  &67\,$\pm$\,4   & 0.05 & 3.818\,$\pm$\,0.080 & 1.01\\
    377443211 & 8500\,$\pm$\,100  & 4.0\,$\pm$\,0.1 & 8300\,$\pm$\,100 & 4.0\,$\pm$\,0.1 &1.7\,$\pm$\,0.2  & 218\,$\pm$\,12& 0.05 & 1.187\,$\pm$\,0.035 & 0.88\\ 
    387757610 & 8200\,$\pm$\,200  & 4.0\,$\pm$\,0.1 & 8300\,$\pm$\,100 & 4.0\,$\pm$\,0.1 & 1.9\,$\pm$\,0.2 & 199\,$\pm$\,9 & 0.02 & 1.365\,$\pm$\,0.038 & 1.51\\ 
    402910080 & 8500\,$\pm$\,200  & 3.5\,$\pm$\,0.1 &                  &                 &                 &270\,$\pm$\,25 & 0.00 & 1.487\,$\pm$\,0.032 & 0.99\\
    424032547 & 14000\,$\pm$\,1000& 3.5\,$\pm$\,0.1 &                  &                 &                 &270\,$\pm$\,27 & 0.02 & 2.724\,$\pm$\,0.063 & 0.88\\ 
    434340780 & 9500\,$\pm$\,100  & 3.5\,$\pm$\,0.2 &                  &                 &                 &270\,$\pm$\,18 & 0.15 & 1.677\,$\pm$\,0.037 & 0.95\\ 
    450085823 & 8500\,$\pm$\,300  & 4.0\,$\pm$\,0.2 &                  &                 &                 &100\,$\pm$\,16 & 0.17 & 1.267\,$\pm$\,0.037 & 0.99\\
    467219052 & 8000\,$\pm$\,100  & 4.0\,$\pm$\,0.1 & 8100\,$\pm$\,100 & 4.0\,$\pm$\,0.1 & 3.1\,$\pm$\,0.1 &111\,$\pm$\,7  & 0.30 & 1.879\,$\pm$\,0.055 & 0.94\\ 
    \hline
\end{tabular}
\end{table*}

\section{Spectroscopic analysis}

\subsection{Radial velocity measurements} \label{sec:rv}

To discover possible variability of the targets induced by orbital motion in a binary system, we used spectra taken at different epochs. There are at least a few hours (mostly a day) or a maximum of around two months between two consecutive spectra of the same star for the same spectrometer. 

\begin{figure*}
\centering
\includegraphics[width=15cm,height=7cm,angle=0]{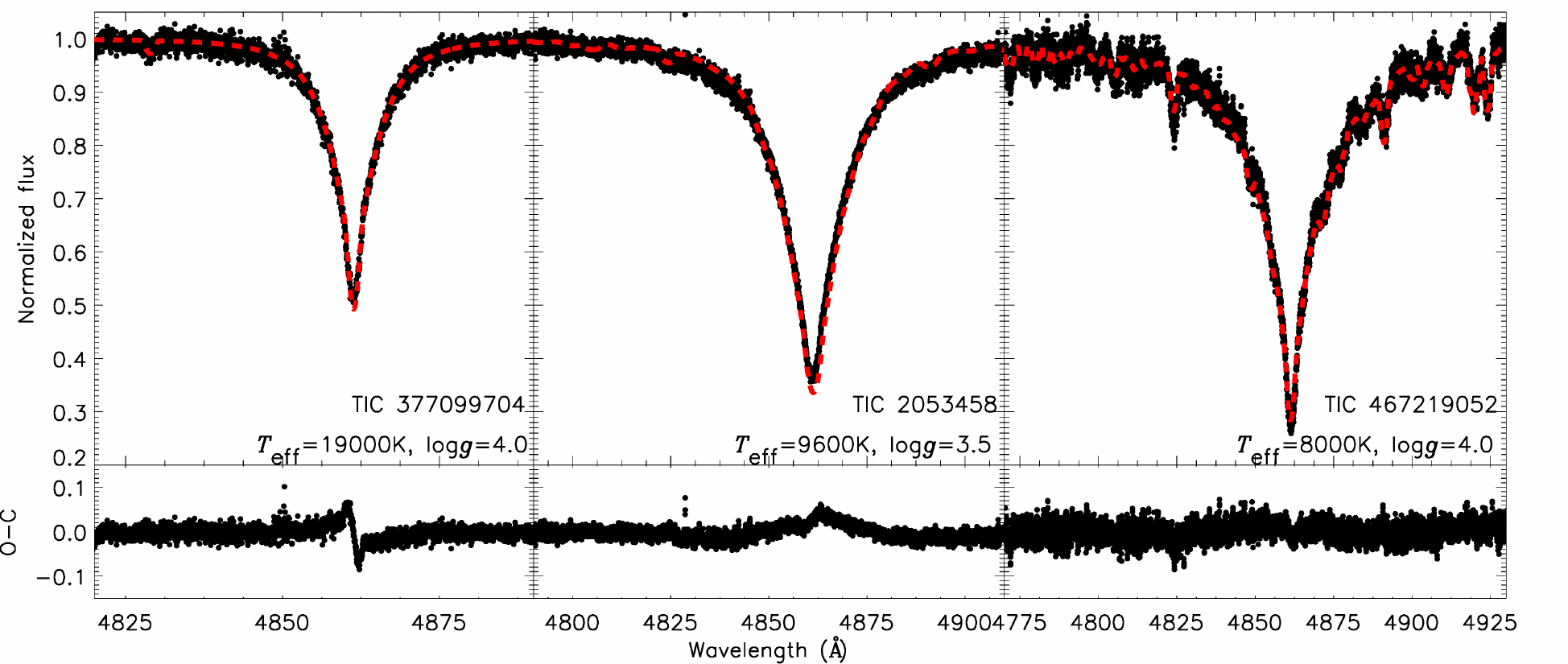}
\caption{Upper panels: Theoretical spectral fits (red dashed lines) to the observed spectra of target stars. Lower panel: Residuals. }\label{fig2_hline_fits}
\end{figure*}

\begin{figure*}
\centering
\includegraphics[width=15cm,height=6cm]{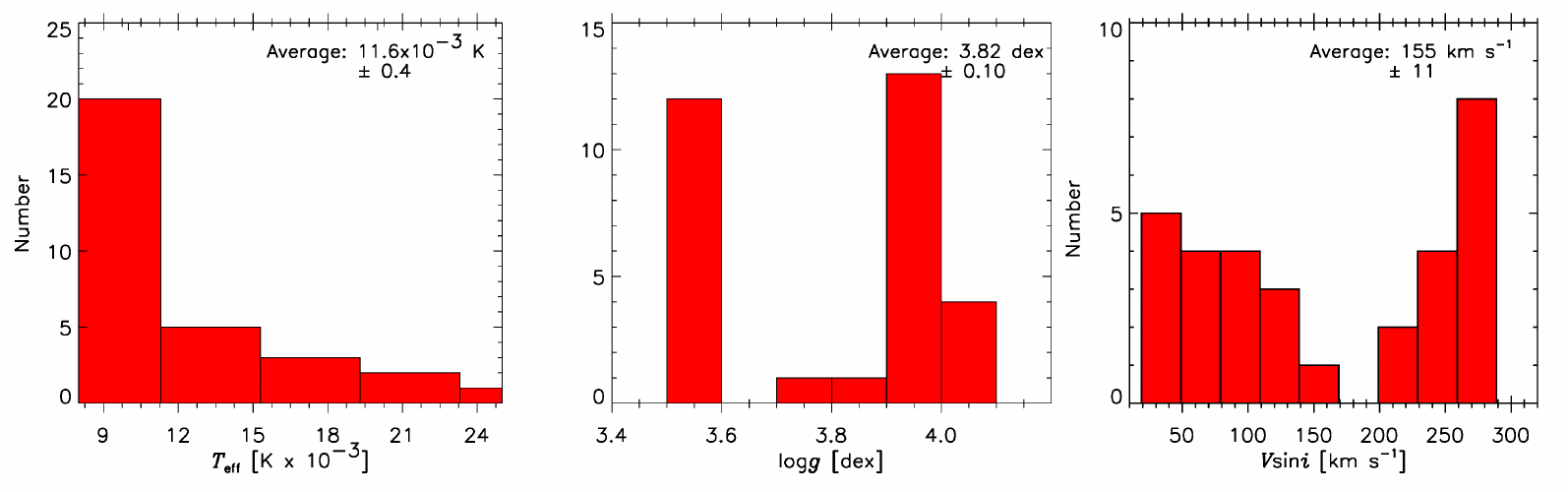}
\caption{Distributions of the \teff, \logg\, and \vsini\, parameters.}\label{para_dist}
\end{figure*}

The radial velocities of the systems were measured using the cross-correlation technique. The IRAF\footnote{http://iraf.noao.edu/}\citep{1986SPIE..627..733T} FXCOR task was used in the investigation. As templates, some synthetic spectra were generated with different \teff\, values of 8000\,K, 10000\,K and 15000\,K.
These atmospheric models were generated using local thermodynamic equilibrium (LTE) ATLAS9 models \citep{1993KurCD..13.....K} and the SYNTHE code \citep{1981SAOSR.391.....K}. Each spectrum of the individual stars was analyzed. However, for most of the stars, no trace of another component and no significant difference between the radial velocity measurements were found within error bars. There are two systems, TIC\,202567458 and TIC\,251196433, whose line profile shapes differ from those of single stars. However, because of high rotation and low SNR values, we could not detect double peaks in cross-correlation analyses of these systems. The line profile shapes of the two atypical systems could be caused by pulsations as well. We classified these systems as suspected binaries. Table\,\ref{rv_table} gives a list of the radial velocity measurements.

\subsection{Atmospheric parameters} 
For the analysis of stars having \teff\, lower than $\sim$15000\,K LTE model atmospheres are convenient \citep{2011JPhCS.328a2015P}. Therefore at the beginning of our study, to estimate the \teff, surface gravity (\logg), and \vsini\, parameters we used LTE ATLAS9 models \citep{1993KurCD..13.....K} and SYNTHE code \citep{1981SAOSR.391.....K} to generate theoretical spectra. The \vsini\, values were determined by the profile fitting method using unblended metal or helium (He) lines \citep{2008oasp.book.....G}. The \teff\, and \logg\, parameters were derived utilizing the hydrogen lines. During the analysis, the minimization method was used to obtain the atmospheric parameters \citep{2004A&A...425..641C}. Since most stars in our list seem to have moderate to high rotational velocity, rather than taking into account the resolving power of the spectrometers we preferred the spectra which have the highest SNR, if a star was observed by both spectrometers. 

According to initially derived \teff\, parameters from the LTE models, some stars were found hotter than 15000\,K. For these stars, the same analysis was carried out by using the non-LTE Tlusty BSTAR2006 grids \citep{2007ApJS..169...83L}. These non-LTE models were generated using the Synspec \citep{2011ascl.soft09022H} program. As a result of LTE and non-LTE analyses, the estimated parameters are listed in Table\,\ref{table2_atmpars}. The uncertainties of the determined parameters were estimated by the 1-$\sigma$ tolerance in the goodness-of-fit parameter. The consistency between the theoretical and observed spectra is shown in Fig.\,\ref{fig2_hline_fits} for some samples.

\subsection{Chemical abundances} \label{chemical}

After we determined the parameters given in Table\,\ref{table2_atmpars}, in the next step we singled out stars that are suitable for chemical abundance analysis. The stars having \vsini\, lower than $\sim$200\,\kms, and spectra with a good SNR were chosen for that purpose. 
During this examination, previously derived atmospheric parameters were taken as input and they were updated utilizing the excitation and ionization equilibrium as described by \cite{2016MNRAS.458.2307K}. 

\begin{figure}
\includegraphics[width=8.5cm,height=4.5cm, angle=0]{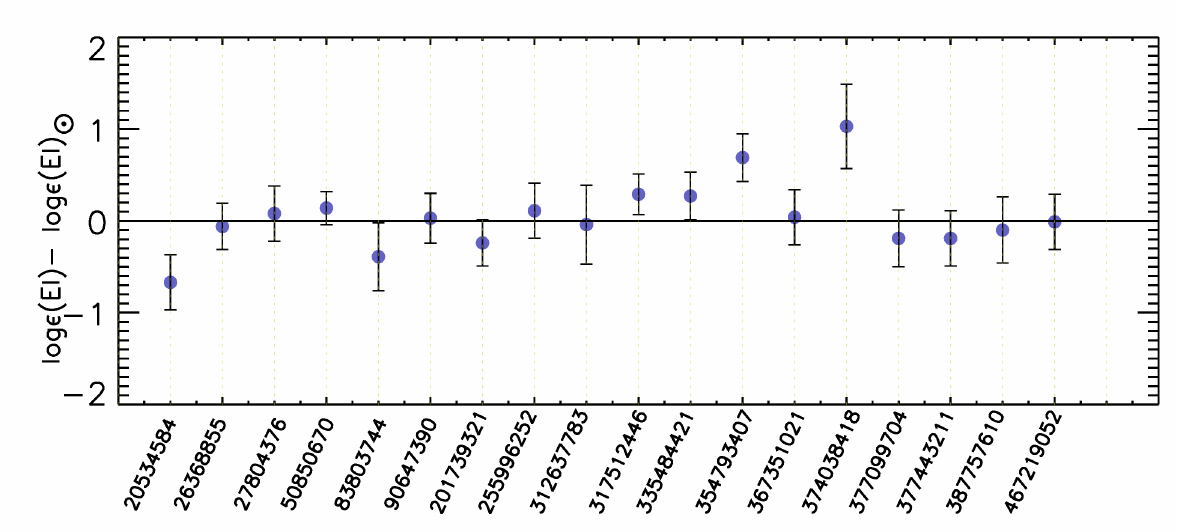}
\caption{The differences between the obtained and solar \citep{2009ARA&A..47..481A} Fe abundances for each star. The x-axis represents the TIC numbers.}
\label{feabundis}
\end{figure}

In the analysis, the Kurucz line list\footnote{http://kurucz.harvard.edu/linelists.html} was used for the line identification and we used the LTE ATLAS9 models.
The spectral synthesis method was considered in the analysis. As a result of the investigation, updated atmospheric parameters and chemical abundances were derived. The list of the updated parameters is given in Table\,\ref{table2_atmpars}, while the element abundances are presented in Table\,\ref{table_abundance}. The distribution of the final  \teff, \logg, and \vsini\, parameters for all targets are shown in Fig.\,\ref{para_dist}. In addition, the Fe abundance distribution of stars is shown in Fig.\,\ref{feabundis}.

\section{Examination of contamination}
\label{sec:tpf_analysis}

\begin{figure*}
    \centering
    \includegraphics[width=\textwidth]{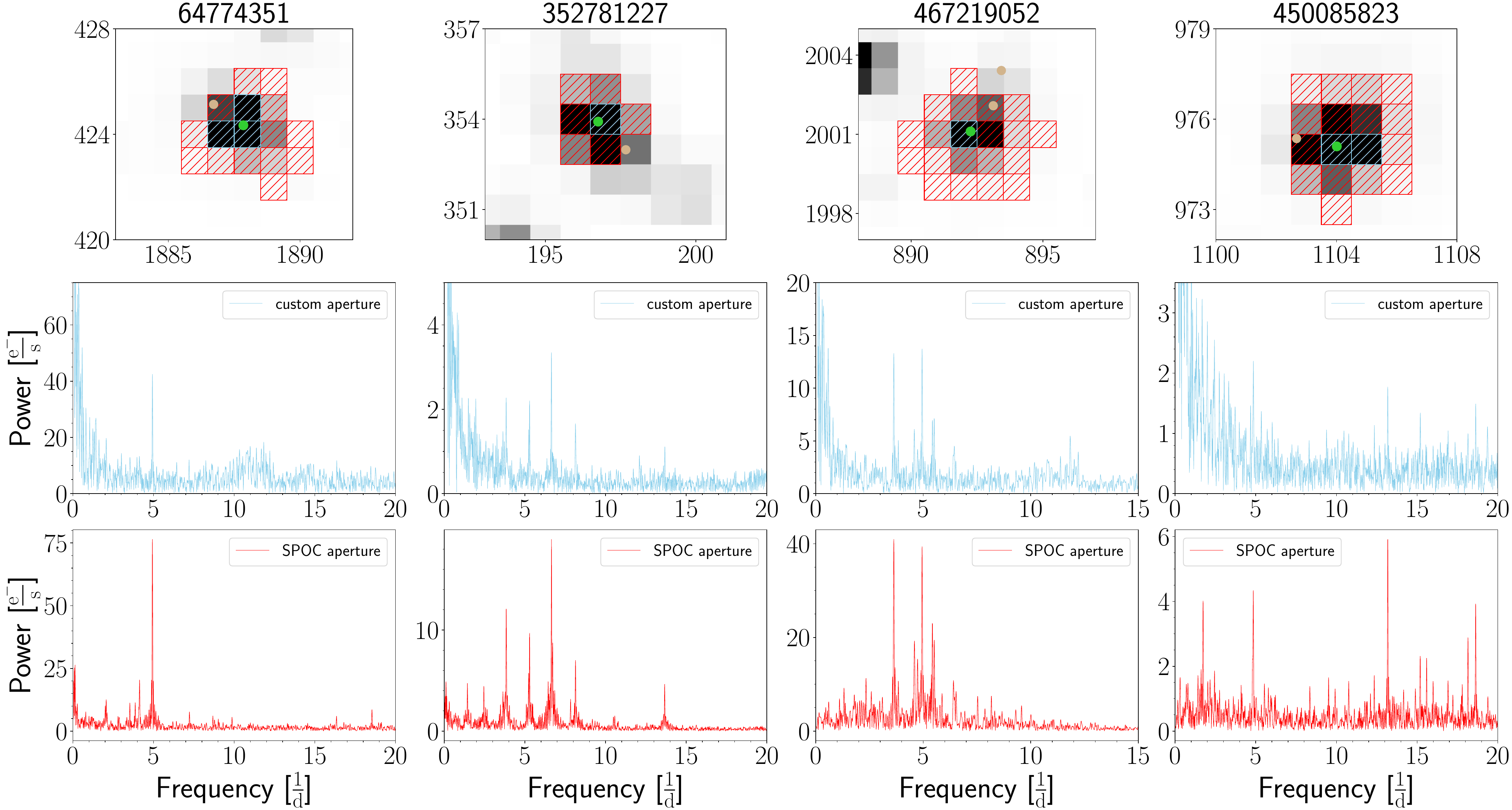}
    \caption{A plot of the {\it {\it TESS}} target pixel files of 4 of the 7 stars discussed in \ref{ap_A}. These stars were the ones for which custom light curves were made. 
    The top panel shows the target pixel files, with the star of interest shown in green (and indicated in the title) and the potentially contaminating stars shown in tan. 
    The middle panel shows the light curve created using a custom aperture (shown in blue in the top panel); the bottom panel shows the light curve generated via the SPOC pipeline. Note that the y-axis scales differ between the middle and bottom panels as a result of the fact that the custom aperture imperfectly captures the flux from the star of interest. This, however, provides the advantage of allowing us to highlight the peak locations.}
    \label{fig:select_custom_lcs}
\end{figure*}

\begin{figure*}
    \centering 
    \includegraphics[width=\textwidth]{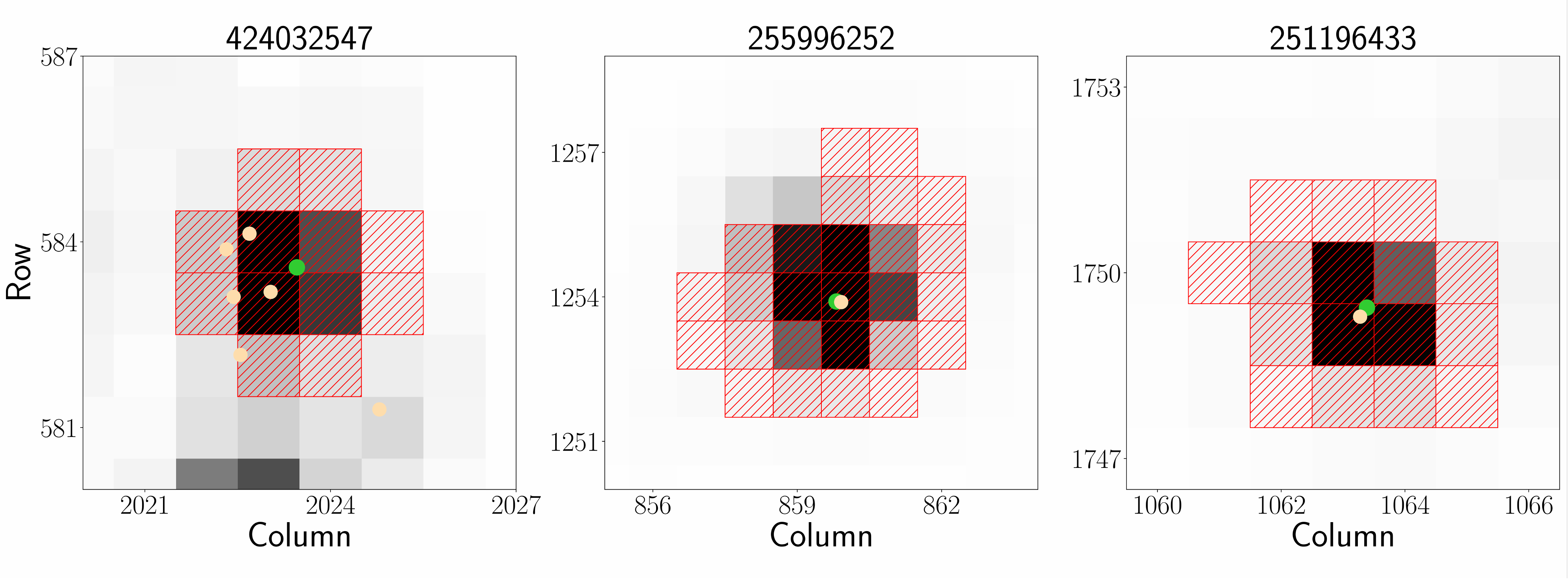}
    \caption{Each panel contains the target pixel file for a star that is so heavily contaminated that we are unable to determine where the true source of the frequency peaks in the periodogram lies. As in Figure \ref{fig:select_custom_lcs}, the green points represent the source of interest, while the tan points represent potentially contaminating stars.}
    \label{fig:other_tpfs}
\end{figure*}

To complement our spectroscopic analyses, we used data from {\it {\it TESS}}. However, the {\it {\it TESS}} pixel scale is 21'', which may induce contamination of the photometric signal from stars located nearby in the sky. To characterize the extent of the contamination in the {\it {\it TESS}} data of our candidates, we worked directly with target pixel files (TPFs).

For any star in which the contamination percentage was greater than 1\%, we queried Gaia DR3 for all the stars brighter than a $G_{\rm mag}$ of 17 within a 200'' radius of the {\it {\it TESS}} source. 
Any targets that had a star within the SPOC photometric aperture that had $\Delta G_{\rm mag}$ < 5 were thoroughly investigated through the use of custom apertures for photometry. 
We checked to see whether we recover the same periodicities for the Maia stars from the SPOC light curve and the light curve calculated using the custom aperture, which will allow us to have confidence in our identification of the pulsation modes.

As part of this analysis, 22 of the 31 stars were found to have contamination ratios greater than 1\%. Several individual stars were selected for individual discussion in \ref{ap_A} if one of the following criteria was true: (i) There are changes in the number or location of significant frequencies when the periodogram of the custom light curve is calculated, (ii) there exist potentially contaminating stars that are within the SPOC photometric aperture and have a $\Delta G_{\rm mag} \leq 3$ compared to the target of interest, or (iii) there is a star with $G_{\rm mag} \leq 14$ whose signal falls into the same pixel as the star of interest. Some of the custom-generated light curves were corrected with cotrending basis vectors to remove long-term trends. The stars that are not discussed in \ref{ap_A} can be considered to have intrinsic variability that is appropriately reflected in the SPOC-generated light curves.

Note that there is a significant tradeoff between the custom light curves and the light curves generated by the SPOC pipeline---while the former will isolate signals to the star of interest, the white noise level will be somewhat higher because the detrending and photometry for this light curve are more rudimentary than the tools SPOC uses. This may mean that some of the lower S/N peaks in the periodogram for the SPOC light curve are washed out by the higher white noise level. Cases where the white noise level of the custom light curve's periodogram is comparable to the height of the peak in the periodogram of the SPOC light curve, are noted below.

Additionally, in some cases, we used the {\it {\it TESS}}-Localize package \citep{2022ascl.soft04005H} to identify the most likely source of some of the frequency peaks in the periodogram. All calls to {\it {\it TESS}}-Localize were used with the {\tt principal\_components} parameter set to 2, in order to de-trend the light curve generated from the TPF.

In Fig.\,\ref{fig:select_custom_lcs}, a plot of the {\it {\it TESS}} pixel files of TIC\,64774351, TIC\,352781227, TIC\,467219052 and TIC\,450085523 is presented. Custom apertures were created containing just the target stars; for some sources, multiple apertures were created to verify the results, as discussed in \ref{ap_A}. The periodograms for the custom apertures are shown in blue in the same figure. According to periodograms and the discussion given in \ref{ap_A}, for TIC\,352781227, TIC\,467219052, and TIC\,450085523 we do not find any significant contamination that would cause frequencies higher than 3\,d$^{-1}$. However, for TIC\,64774351 the result of contamination analysis could not give a reliable result for high-frequencies seen in the original periodogram. 


In Fig.\,\ref{fig:other_tpfs}, the heavily contaminated stars are shown. For TIC\,255996252 according to our investigation, we found that the target of interest is the original source of high-frequencies (for detail see \ref{ap_A}). However, for other systems, we were unable to distinguish whether or not the high-frequency peaks come from our targets. 

\section{Discussion}\label{discussion}

\begin{figure}
\centering
\includegraphics[width=8.7cm,height=6.0cm, angle=0]{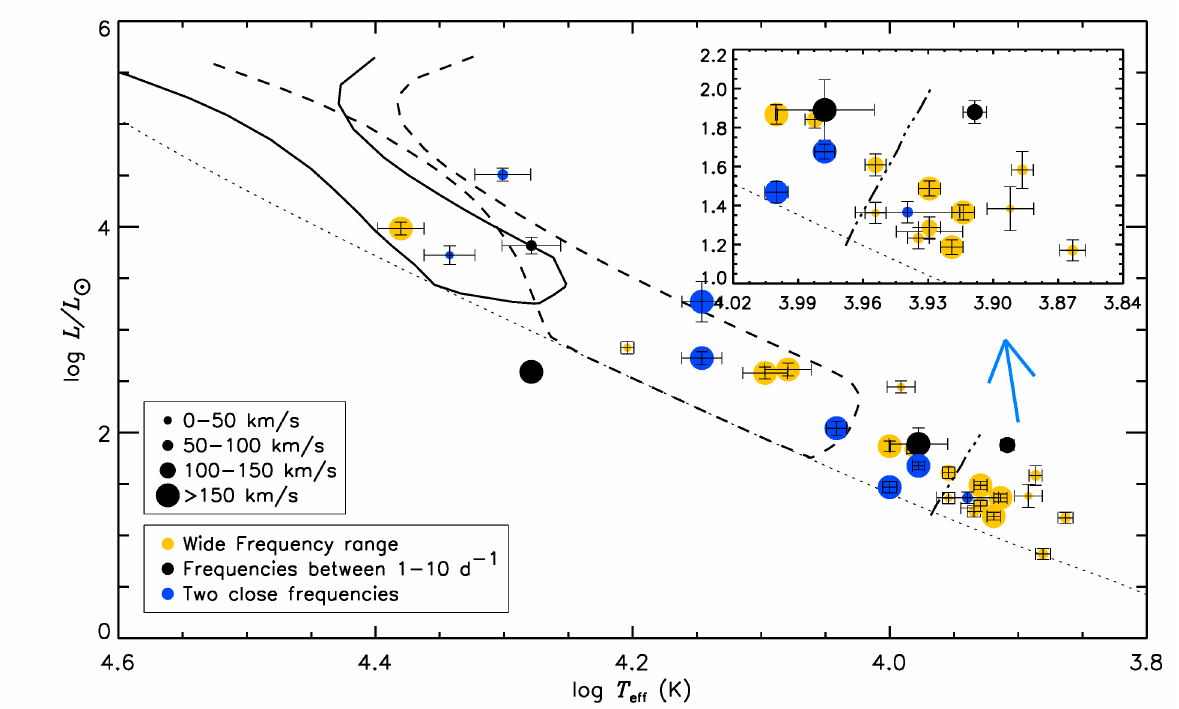}
\caption{The final positions of the selected Maia candidates in the H-R diagram. The theoretical instability strips of \bc\,, SPB and \ds\, stars were taken the same as in Fig.\,\ref{figure1_first_position}. A detailed explanation of the brief expressions of frequencies in the figure can be found in Sect.\,\ref{discussion}. Some targets are not included in this figure, since they were excluded from the Maia candidate list according to the results of our contamination analysis. }
\label{final_position}
\end{figure}

We now proceed to the overall discussion of the variability of our sample of target stars. Each individual star is considered in more detail in \ref{ap_B}.
We present the positions of our targets in the theoretical H-R Diagram in Fig. \ref{final_position}. The frequency spectra of all systems are shown in Fig\,\ref{specall} in order from the hottest star to the coldest star.

The positions of the candidates in this diagram take into account the final atmospheric parameters, the structure of frequency spectra, and the \vsini\, values. 
To plot the final H-R diagram the $L$ parameters were recalculated taking into account the final \teff\, values. Additionally, the E(B-V) parameters were re-determined from the interstellar sodium lines as described in the study of \cite{2016MNRAS.458.2307K} (cf. Table \ref{table2_atmpars}).

\subsection{Incorrect \teff\, hypothesis}

\begin{figure}
\includegraphics[width=8.5cm,height=6.5cm, angle=0]{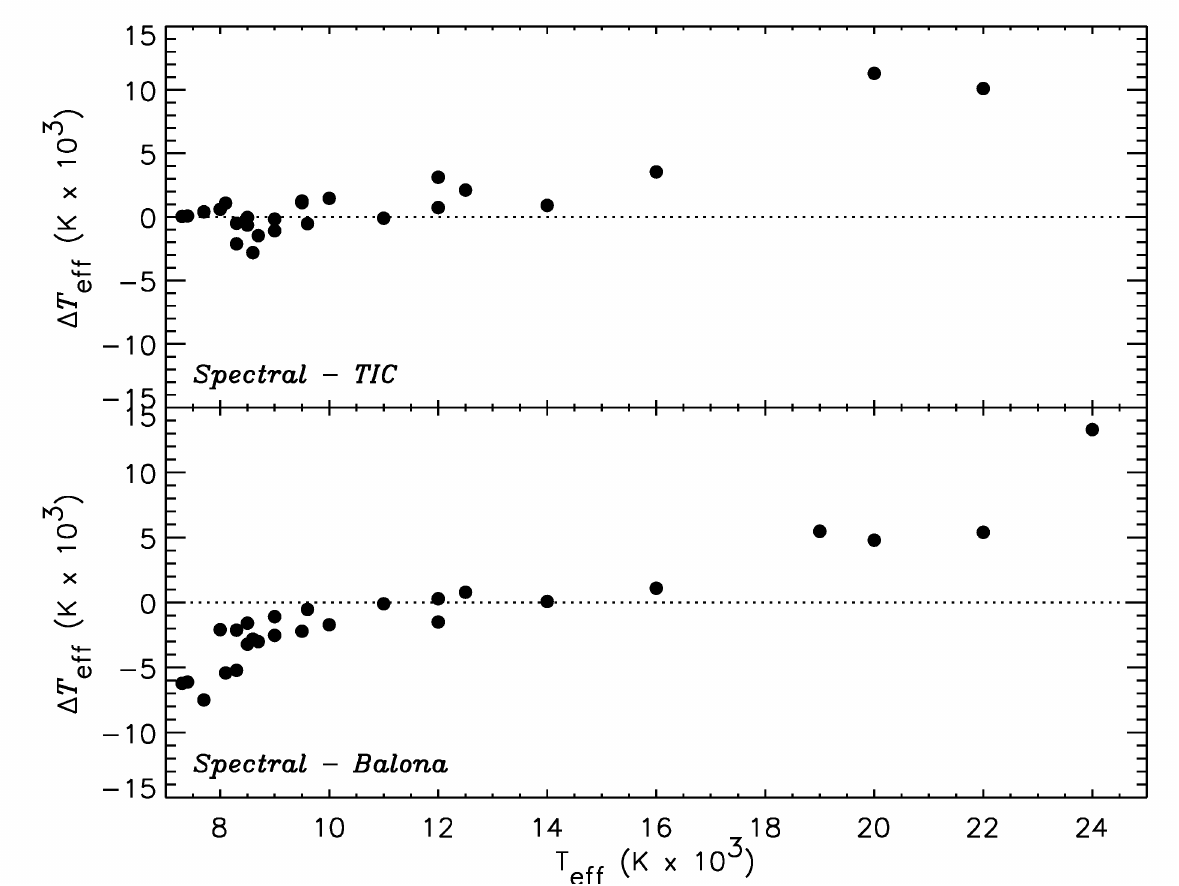}
\caption{Comparison of the \teff\, of the TIC \citep{2019AJ....158..138S} (upper panel) and \citet{2020MNRAS.493.5871B} (lower panel) with the spectral \teff\, determined in this study. The x-axis shows the spectral \teff\, values while the y-axis illustrates the \teff\, differences between the spectral and TIC (upper panel) and \citet{2020MNRAS.493.5871B} (lower panel).}
\label{teff_comp}
\end{figure}

In stark contrast to Fig.\,\ref{figure1_first_position} most of the target stars now lie in known pulsational instability regions, suggesting that most literature \teff\ values used to select them are inaccurate. We compared our spectral \teff\, values with the \teff\, of the TIC and \cite{2020MNRAS.493.5871B} in Fig.\,\ref{teff_comp}. As can be seen from the figure there are significant differences between the \teff\, values. This result is an expected conclusion for the TIC \teff\, values for hot stars because of the uncertain reddening corrections for stars in the Galactic plane. A similar explanation may apply to the \teff\, values used by \cite{2020MNRAS.493.5871B}, as those were derived mostly from photometric colours. According to this comparison, we conclude that for most targets previously derived \teff\, values show significant discrepancies from the spectral ones. Even so, in Fig. \ref{final_position} there are still some stars located outside of any known pulsational instability strip.

At this point it is important to stress that the instability strip boundaries depicted in Figs. \ref{figure1_first_position} and \ref{final_position} are not immutable. The theoretical boundaries ($\beta$ Cep stars, SPB stars) depend on the input physics used, as discussed by \cite[e.g.,][]{2023arXiv231108453A}, and the observational boundaries for the $\delta$ Scuti stars indicate where pulsations are less likely \citep{2019MNRAS.485.2380M}. Nevertheless, these Figures give an idea in which stars pulsations could be difficult to explain.

\subsection{Binary hypothesis}

According to the binary hypothesis, it was suggested that late B/early A type pressure mode pulsators could be binary systems including a B-type star with a \ds\, companion. Hence, the system seems to be an oscillating star but it is too cool to be a \bc\, and too hot to be a \ds\, star.
As a result of our investigation of the spectral line profiles and radial velocities of the stars in our sample, we did not reveal any obvious binary systems. Only two systems were classified as suspected binary systems because of their line profile changes, which however could be well caused by pulsations. Additionally, our method is not sensitive to binary systems with long orbital periods (more than a few months) given the temporal sampling of our spectra. Also, many stars in our sample show high rotational velocity (>100\,\kms) and some spectra have low SNR, which affects the radial velocity measurements and the detection of a possible binary component in the systems. Consequently, our spectroscopic investigation of possible binarity remained mostly inconclusive.

Another possibility to detect binaries would be to search for phase modulations of the pulsational signals \citep{2014MNRAS.441.2515M}. Unfortunately, we were not able to find any evidence thereof, presumably owing to the small amplitudes of the pulsations detected and the relatively short duration of the individual {\it TESS} light curves.

We also examined the Gaia Renormalized Unit Weight Error (RUWE \citeauthor{2021A&A...649A...2L} \citeyear{2021A&A...649A...2L}). RUWE is a parameter that shows the degeneration in the astrometric parameters determined with Gaia \citep{2021A&A...649A...6G}. High values of RUWE (higher than 1.4) indicate that the system of interest could be a binary \citep{2018A&A...616A...2L, 2021A&A...649A...2L}. \cite{2021ApJ...907L..33S} also point out that any RUWE value above 1.0 may already be indicative of binarity. Therefore, we collected the RUWE parameters of our targets from Gaia DR3 \citep{2021A&A...649A...6G} which are given in Table\,\ref{table2_atmpars}. As can be seen from the table, there are nine systems that have RUWE\,>\,1.4. The largest values, 11.59 and 20.59, relate to TIC\,36557487, and TIC\,367351021, respectively. One suspected binary system from our radial velocity analysis (TIC\,251196433) has a RUWE value of 2.01. The systems with high RUWE values (>\,1.4) are likely astrometric binaries, 
even if none of our targets is in the list of non-single stars according to Gaia DR3. Catalogues of double and/or multiple stars were also searched \citep[e.g.][]{2012AstBu..67...44B,2008MNRAS.389..869E,2002A&A...384..180F, 2001AJ....122.3466M,1997ESASP1200.....E, 1995IAUS..166..395D}. Indeed some of our targets were found in these catalogs and are classified as double or multiple systems. TIC\,36557487, TIC\,64774351, and TIC\,354793407 are members of visual double systems. 

\subsection{Rapid rotation hypothesis}

As discussed in the introduction, rapid rotation could a) modify a star's position in the H-R Diagram, b) shift the observed pulsation frequencies, c) excite Rossby waves, and d) provide means of $\kappa-\gamma$-driven excitation of p and g modes. To evaluate our candidate stars with respect to rotation we first took into account the stars which show frequencies between 1-10 d$^{-1}$. There are four stars that show frequencies in that range (TIC\,251196433, TIC\,352781227, TIC\,377099704, TIC\,467219052) that are shown with black dots in Fig.\,\ref{final_position} and two of them exhibit high rotation; TIC\,251196433 and TIC\,352781227. 
TIC\,251196433 is positioned slightly below the instability strip, situated between SPB and \bc\, stars. This fast-rotating star  ($\gtrsim$255\,\kms) could either be a hybrid pulsator or an SPB star displaying shifted pulsation frequencies. TIC\,352781227 is also a fast-rotating star ($\gtrsim$270\,\kms) and placed between the \ds\, and SPB instability strip. For this star, high rotational velocity could cause low frequencies to be shifted towards high frequencies and since we likely see this star close to equator-on, our \teff\, determination would be too low.
Therefore, this variable is likely an SPB star. A more detailed discussion on this object is also given in \ref{ap_B}. 

The other stars showing frequencies between 1-10 d$^{-1}$ have low to moderate \vsini. One (TIC\,467219052) is simply a somewhat evolved \ds\, star, while the other is located in the SPB domain, very close to the \bc\, territory (TIC\,377099704). These stars are hence well placed in known instability strips. Both show both low and high frequencies, these stars could be SPB\,$-$\,\bc\, and a \ds\,$-$\gd\, hybrids. 

There is a group of stars that are distinguished by two close frequencies dominant in the Fourier spectra (TIC\,26369887, TIC\,36557487, TIC\,312637783, TIC\,374038418, TIC\,434340780). As depicted in Fig.\,\ref{final_position} most of these stars exhibit rapid rotation with some falling between the SPB and \ds\, domains. Because of high rotational velocity, the rapidly rotating ones could appear cooler than their actual temperature, leading to a shift in their pulsation frequencies towards higher values. We therefore classify them as cool SPB stars. Further discussions on these targets are provided in \ref{ap_B}.

When examining other stars exhibiting oscillations in a wide frequency range (up to 50\,d$^{-1}$), we noticed that most of them are located inside the \ds\, instability strip. However, six of them are placed beyond the hot border of \ds\, instability strip. Among these, TIC\,20534584 lies between the \ds\, and SPB instability strips. Notably, it possesses a moderate to high \vsini\, (150\,\kms), and its relatively high RUWE parameter (2.26) suggests a potential binary nature with a hotter companion, placing it beyond the hot border of \ds\, instability strip. 
There are also four stars (TIC\,83803744, TIC\,192780911, TIC\,236785664, TIC\,354793407) placed inside the SPB domain and showing frequencies up to 40 d$^{-1}$. While two of them have low \vsini\, values, TIC\,83803744 exhibits a low RUWE parameter, and TIC\,354793407 has a RUWE of 1.81. 
Regarding TIC\,83803744 its high frequencies lack a clear explanation. However, it is classified as HgMn star (see \ref{ap_B}). As HgMn stars are often found in binaries (e.g., \citealt{1985A&A...146..341G}), which suggests a binary nature. 
The remaining two systems (TIC\,192780911 and TIC\,236785664) exhibit high rotational velocities with high values of the RUWE parameter indicative of a possible binary nature. TIC\,192780911 displays a weak signal at around 3.9\,$^{-1}$, possibly indicating an SPB-type pulsation modified by the rapid rotation. 
TIC\,236785664 shows both low and high frequencies typically of \ds\, stars, yet it's \teff\, exceeds the typical range for \ds\, stars. 
Finally, TIC\,202567458, a rapid rotator, manifests a pulsation spectrum comprising apparent g-mode pulsations, their linear combinations, and a stochastic component implying the presence of Internal Gravity Waves (IGWs, e.g., see \citealt{2023A&A...674A.134R, 2019NatAs...3..760B}). The higher frequency signals are compatible with combination frequencies.

\subsection{Variability classification of the target stars}

We summarize the discussions above and the results on the individual stars presented in \ref{ap_B} in Table \ref{tab:varclass}, respectively. Out of the 31 stars considered in this paper and previously claimed as variables outside of any known instability strip, the largest group (14 objects) are normal \ds\,stars that simply had overestimated \teff. Six stars are gravity mode pulsators with combination frequencies that protrude into the pressure mode domain, but these higher frequencies are not due to independent pulsation modes \citep[cf.][]{2015MNRAS.450.3015K}. Among the hottest stars, we found one star with gravity modes and Internal Gravity waves, two \bc\, pulsators, and two \bc\-/SPB "hybrids".

\begin{table}
    \centering
\caption{Variability classification of our target stars.}
 \label{tab:varclass}
    \begin{tabular}{ll}
    \hline
    TIC       & Variable type	\\
\hline
16781787 & SPB	\\
20534584 & g modes + comb. + \ds	\\
26368855 & \ds\	\\
26369887 & SPB + combinations	\\
27804376 & \ds\	\\
36557487 & g modes + combinations	\\
50850670 & \ds\ + $\gamma$ Dor	\\
64774351 & g modes + combinations	\\
83803744 & ROT + \ds\,binary?	\\
90647390 & \ds\ + binary?	\\
192780911 & SPB + binary?	\\
201739321 & \ds\, + ROT?	\\
202567458 & g modes + IGW	\\
236785664 & Be/SPB + \ds\,?	\\
251196433 & "hybrid" \bc\,/SPB 	\\
255996252 & \ds\	\\
312637783 & \bc\, + binary?	\\
317512446 & \ds\	\\
335484421 & \ds\	\\
352781227 & g modes + combinations	\\
354793407 & ROT + \ds\	\\
367351021 & \ds\ + binary?	\\
374038418 & \bc\, + IGW	\\
377099704 & "hybrid" \bc\,/SPB 	\\
377443211 & \ds\	\\
387757610 & \ds\ + $\gamma$ Dor	\\
402910080 & \ds\	\\
424032547 & contamination	\\
434340780 & g modes + combinations	\\
450085823 & \ds\	\\
467219052 & \ds\	\\
    \hline
    \end{tabular}
\end{table}

Regarding the remaining six objects, the {\it TESS} light curve of one target was revealed to be contaminated by a background ellipsoidal variable. Two stars were identified as SPB stars with a very fast rotation that presumably modifies the pulsation frequencies in the observer's reference frame to values that are comparable with those of pressure modes. That leaves three stars, two of which are in the SPB star instability strip and one is in between the SPB and the \ds\ star domains. Two of these stars are rotational variables with additional frequencies in the \ds\ range. For both of those, there is some indirect evidence of binarity, one TIC 354793407 = HD 21699 has a RUWE of 1.81, and the other TIC 83803744 = HD 1009 is a HgMn star which in most cases are binaries. Consequently, the short-period signals in these star's light curves might well arise from a \ds\ companion. The last object is a rapidly rotating Be star with a somewhat high RUWE and several frequencies typical for \ds\ stars. This latter object, TIC 236785664 = HD 175511, deserves some further investigation.

\section{Conclusions}

In this study, a detailed analysis of 31 stars that have been claimed as late B/early A type short period pulsators is presented. In this part of the H-R diagram no such pulsations are expected, so these objects need to be explained. There are four different hypotheses that are to be considered: incorrect \teff, binarity, rapid rotation and possible contamination of the light curves. To this end, a detailed spectroscopic investigation was carried out for our candidate variables. For all stars accurate stellar parameters such as \teff, \logg, \vsini\, were determined. Furthermore, individual element abundances were derived for all stars where this was possible. In addition, a contamination analysis was carried out to be sure that the determined high pulsation frequencies arise from the stars of interest. 

There are several stars in our sample whose frequency spectra appear phenomenologically similar, hence we analysed them in groups. The first group shows a wide frequency range extending to $\sim$40\,d$^{-1}$. Another group exhibits two dominant frequencies close to each other and the final group possesses frequencies between $1-10$ d$^{-1}$. Taking into account these groups, the positions of the stars in the H-R diagram, \vsini\, values, the radial velocity measurements, and the RUWE parameter we discussed the proposed hypotheses of tthe existence of the so-called "Maia variables" located on the main-sequence in between the instability strips of the SPB and $\delta$\,Scuti stars. First, we conclude that incorrect \teff\, values by themselves can not explain all of these stars even if for some systems there are significant differences between the spectroscopic and photometric \teff\, values (see Fig.\,\ref{teff_comp}). The binary hypothesis alone is also insufficient. When we examined the high rotation hypothesis by considering the systems' positions in the H-R diagram and the structure of frequency spectra. In this way, we identified some stars whose frequencies were likely shifted towards higher values due to rapid rotation.

However, taking into account a combination of all hypotheses, most of the 31 stars we studied can be explained easily. 27 are members of established variability classes, ranging from Internal Gravity Wave pulsators, \bc\ stars, SPB stars to \ds\ and $\gamma$ Dor stars. The variability of one object is due to a contaminating neighbouring star. There are three stars that are not that easy to explain. Two of them are hot rotational variables which also exhibit short-period pulsations, but for both of them, we found indirect evidence for binarity. The last star is a rapidly rotating Be star in the SPB instability strip that also shows multiperiodic short-period pulsations, but also, in this case, the Gaia RUWE parameter is suspicious in terms of binarity.

Extrapolating our results to the hundreds of main sequence pulsators of B/A spectral type outside of any known instability region claimed recently \citep{2023FrASS..1066750B}, we expect that most if not all of these objects can be explained in different ways as well. From today's point of view, there is thus no need to claim the existence of an unexplained class of variable stars in this domain of the H-R diagram, let alone call them "Maia variables" as Maia itself is not a pulsating star \citep{2017MNRAS.471.2882W}.


\begin{acknowledgement}
The authors would like to thank the reviewer for useful comments and suggestions that helped to improve the publication. This study has been supported by the Scientific and  Technological  Research Council (TUBITAK) project 121F474. GH thanks the Polish National Center for Science (NCN) for supporting the study through grants 2015/18/A/ST9/00578 and 2021/43/B/ST9/02972. Horizon 2020: (OPTICON) This project has received funding from the European Union's Horizon 2020 research and innovation programme under grant agreement No 730890. The calculations have been partly carried out using resources provided by the Wroc\l{}aw Centre for Networking and Supercomputing (http://www.wcss.pl), Grant No 214. This material reflects only the authors views and the Commission is not liable for any use that may be made of the information contained therein. Based on observations made with the Mercator  Telescope,  operated on the island of La  Palma by the Flemish Community, at the Spanish Observatorio del Roque de los Muchachos of the Instituto de Astrof\`{\i}sica de Canarias. The {\it TESS} data presented in this paper were obtained from the Mikulski Archive for Space Telescopes (MAST). Funding for the {\it TESS} mission is provided by the NASA Explorer Program. This work has made use of data from the European Space Agency (ESA) mission Gaia (http://www.cosmos.esa.int/gaia), processed by the Gaia Data Processing and Analysis Consortium (DPAC, http://www.cosmos.esa.int/web/gaia/dpac/consortium). Funding for the DPAC has been provided by national institutions, in particular, the institutions participating in the Gaia Multilateral Agreement. This research has made use of the SIMBAD database, operated at CDS, Strasbourg, France. We are grateful to Vichi Antoci, Tim Bedding, Dominic Bowman, Don Kurtz, Nami Mowlavi and Simon Murphy for their comments on a draft version of this paper.
\end{acknowledgement}

\bibliography{pasa_Maia}

\appendix

\renewcommand{\thefigure}{A\arabic{figure}}
\setcounter{figure}{0}  

\begin{figure*}
 \begin{minipage}[b]{0.98\textwidth}
  \includegraphics[angle=270, width=0.992\textwidth]{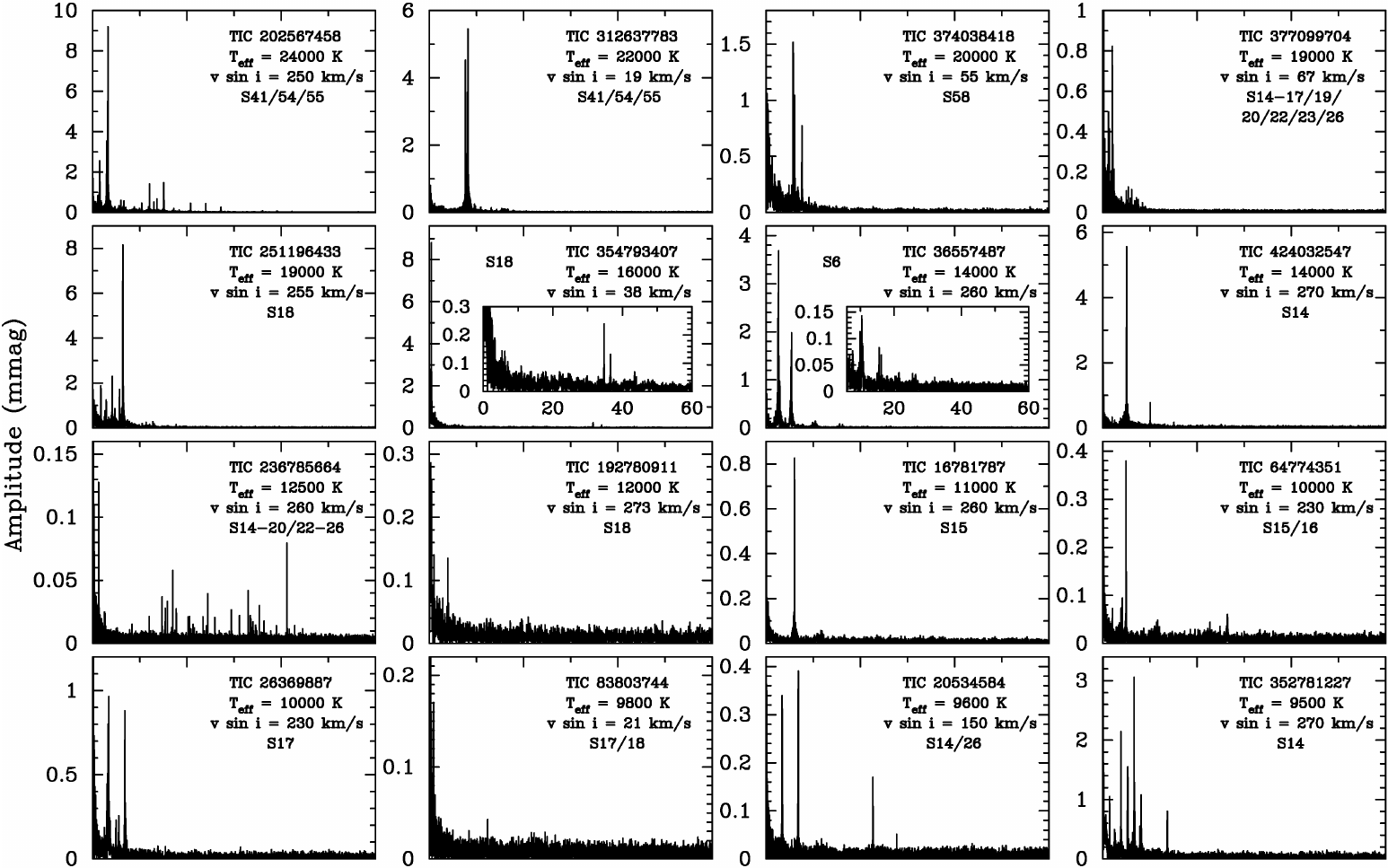}
  \end{minipage}
  \begin{minipage}[b]{0.98\textwidth}
  \includegraphics[angle=270,width=1\textwidth]{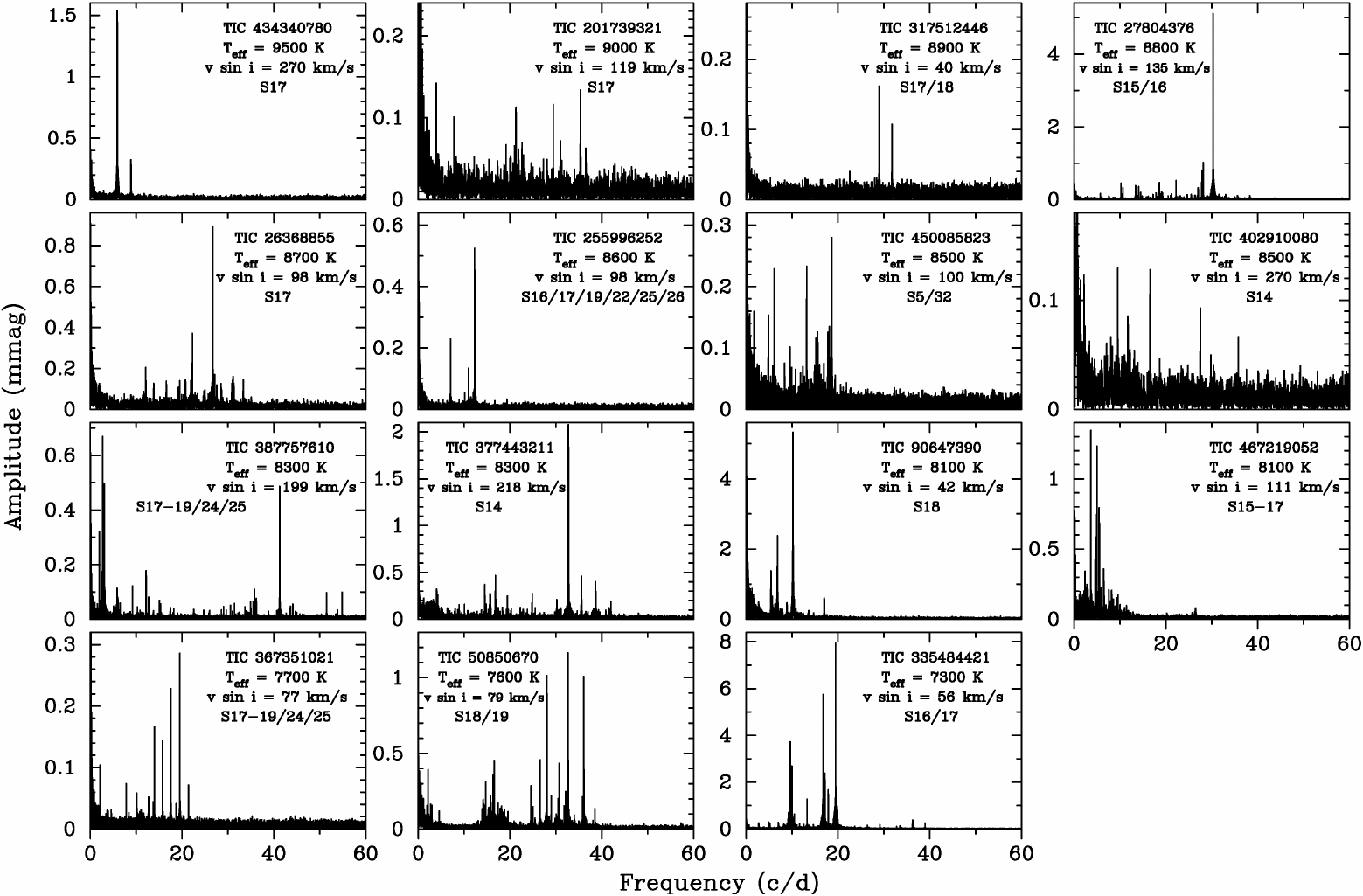}
  \end{minipage}
\caption{Amplitude spectra of the targets ordered by decreasing \teff. The {\it TESS} data sectors used for those plots are indicated.}\label{specall}
\end{figure*}

\section{Contamination analysis of individual stars} \label{ap_A}
 
\subsection{TIC 64774351}

The SPOC-derived light curve for TIC\,64774351 has only 90.3\% of its flux arising from the star of interest. 
There are two stars with $9 \leq G_{\rm mag} \leq 10$ within 100'' of TIC\,64774351 (BD+52 2951 and BD+52 2952); the former is especially a key potential source of contamination. 
A light curve created from an aperture that does not include these two stars still shows a highly significant peak at 0.203\,d, as shown in the left panel of Figure \ref{fig:select_custom_lcs}. Another choice of aperture,
centered on the pixels to the right of the star of interest, provides some evidence for the peak at $\sim$\,5\,d$^{-1}$ being associated with this star. However, it is unable to resolve the forest of peaks
at higher frequencies (e.g., those around 10--20\,d$^{-1}$) neither can the periodogram generated from
the custom light curve for the aperture shown in Figure \ref{fig:select_custom_lcs}.
However, the white noise level is higher in the periodogram created from the custom aperture, which could hide these peaks, which are already relatively low signal-to-noise in the SPOC light curve. 
The nearby star, BD+52\,2951 (indicated in tan in the top left panel of Figure \ref{fig:select_custom_lcs}), has a listed spectral type of F8, suggesting that it is likely not in the classical instability strip and should thus not pulsate.

An analysis of this star using the {\it TESS}-Localize tool suggests that the lower frequencies (those less than $\sim 5$d$^{-1}$) are highly likely to be associated with BD+52\,2952. However, the analysis was unreliable for the higher frequencies (i.e., those greater than $\sim 5$\,d$^{-1}$), so in this case, we are unable to use {\it TESS}-Localize to draw any conclusions.

\subsection{TIC 251196433}

The SPOC-derived light curve for this star has only 94.1\% of its flux arising from the star of interest. There exists a star with $G_{\rm mag} = 14$ within 4'' (Gaia DR3 460981833479443456; henceforth referred to as GDR3-456); their locations in the TPF are shown in Figure \ref{fig:other_tpfs}. These two stars fall within the same pixel, making it nearly impossible to disentangle their respective variability via the use of custom apertures. \cite{2020AJ....160...32L} identified TIC\,251196433 as a $\beta$ Cep star based on its low-frequency variability (most of the peaks are at frequencies less than 10 d$^{-1}$). However, this is also the regime of certain compact g~mode pulsators.

Given that GDR3-456 has a similar parallax to TIC\,251196433, they likely lie at similar distances, and so we can assume that the extinction coefficients (and thus, color excess) for both stars are roughly identical. Using the Gaia value of E($G_{\rm BP} - G_{\rm RP}$) for TIC\,251196433, we see that the color of GDR3-456 is around -0.17, which could make it a cool sdB star. The pulsation amplitude is 0.7\% for the highest amplitude frequency in the periodogram, which is too high for an sdB g~mode pulsator and is thus likely associated with TIC\,251196433. However, the other peaks in the periodogram represent pulsation amplitudes of 0.15\% or less; these modes could arise from the contaminating star GDR3-456.\footnote{For an example of sdB stars pulsating in this regime, see \citet{2019A&A...632A..90C}.} Further analyses, including speckle imaging and ground-based spectroscopy, may be needed to tease out the properties of this star and separate its modes from TIC\,251196433.

\subsection{TIC 255996252}

The SPOC light curve for this star has only 90.8\% of its flux arising from the star of interest. As for TIC\,251196433, and as shown in Figure \ref{fig:other_tpfs}, this star is nearly coincident in the sky with another star, Gaia DR3 2274443263522286208 ($G_{\rm mag}$ = 12.49; henceforth GDR3-208). This star is located at a similar distance as TIC\,255996252, at around 210 pc based on parallax. GDR3-208, however, does not have any information about its color from Gaia DR3 or prior survey missions such as Hipparcos. Assuming similar extinctions $A_G$ (a value is provided for this for TIC\,255996252 in Gaia DR3), we get an absolute magnitude in $G$-band of 5.4, suggesting that it is likely a G-type star. Given that such stars' pulsation modes are convectively driven and exhibit clear frequency spacings, this star likely does not contaminate the light curve of TIC\,255996252, and all the peaks seen in the star's Fourier spectrum can be taken to be intrinsic to this star.

\subsection{TIC 352781227}
The SPOC light curve for this star has only 84.5\% of the flux arising from TIC\,352781227. This star has $G_{\rm mag} = 11.28$, and there is a nearby star (within 30'') with $G_{\rm mag} = 12.82$. We created a custom aperture for the target pixel file containing just this star (shown in light blue shading in Figure \ref{fig:select_custom_lcs}, compared to the SPOC aperture in red shading). 
While this of course does not contain the entirety of the flux from this star (and so the peak amplitudes in the periodogram will be correspondingly lower), the periodogram should capture the frequencies of the peaks intrinsic to this star's variability. We find that the noise level of the periodogram is too high in the low-frequency regime ($\nu \lesssim 3$\,d$^{-1}$), so no conclusions can be drawn about these frequencies; however, the higher-frequency ones ($\nu \geq$\,3\,d$^{-1}$) are very likely intrinsic to the star.

Another aperture that includes only the two topmost pixels in the SPOC aperture exhibits a peak
at 6.5\,d$^{-1}$, suggesting that it is intrinsic to the star. No evidence for higher frequencies
is present, and the noise level at low frequencies is too low to draw any meaningful conclusions.


\subsection{TIC 424032547}\label{tic4240}
Of all the stars in our sample, this star has the highest likelihood for contamination, as only 72.7\% of the flux in the aperture comes from the star of interest. There is one significant peak at 5.03 d$^{-1}$, with at least 5 significant harmonics in the periodogram. An analysis of this TPF with {\it TESS}-Localize suggests that the probability that this frequency (and, by extension, its harmonics) is associated with that particular star is quite low. Indeed, as noted in \cite{2020AJ....160...32L}, this star is located in a dense region of an open cluster, such that the light curve and target pixel file suffer from significant contamination and saturation that could affect the SPOC light curve.

Additional verification that the observed signal does not arise from this particular star lies in the {\tt VARIABLE} flag assigned to stars in the Gaia catalog. There are two nearby stars (designated Gaia DR3 1823376830245623296 and DR3 1823376825919999232), which are somewhat fainter; however, both are designated with the {\tt VARIABLE} flag in Gaia, while TIC\,424032547 is not (its value for this flag is {\tt NOT\_AVAILABLE}). While this is not a sufficient condition to say that the observed variability does not arise from TIC\,424032547, it is further strong evidence that suggests that the SPOC light curve does not capture the intrinsic variability for this star. Additionally, an aperture with the rightmost 4 pixels from the SPOC aperture does not show
any evidence of variability at the observed frequencies, providing further evidence that the observed
signals in the periodogram do not arise from the star of interest.

The nearby contaminants are shown in the left panel of Figure \ref{fig:other_tpfs}.

\subsection{TIC 450085823}
There is a nearby star with G$_{\rm mag}= 14.04$, as shown in the rightmost panel of Figure \ref{fig:select_custom_lcs}. This star contributes 2.73\% of the flux to the aperture used to construct the SPOC light curve, so it will likely have a minimal impact on the frequencies recovered from the light curve. We constructed a custom light curve using a two-pixel aperture and recovered most of the large amplitude peaks visible in the SPOC light curve, albeit at a lower S/N ratio. The peak at $\nu\sim 1.8$\,d$^{-1}$ is not visible in the custom light curve, perhaps due to the fact that it arises from the contaminating star. However, it could also simply be hidden below the noise. The high-frequency peaks are still visible in the custom light curve's analysis.

\subsection{TIC 467219052}
This star has two potential nearby contaminants with Gaia magnitudes 12.06 and 12.94 (as seen in the third panel from left in Figure \ref{fig:select_custom_lcs}); the SPOC light curve captures 88\% of the flux from the star of interest. A custom light curve from the single-pixel aperture shows that the largest peaks below 7\,d$^{-1}$ are likely intrinsic to this star. The peak at $\sim$11.8\,d$^{-1}$ is significant in the custom light curve but is not present in the SPOC light curve. This may arise from some systematic variability in the custom light curve that was unable to be removed using the CBVs. Selecting the 6 leftmost pixels in the SPOC aperture
and constructing a light curve yields a periodogram with significant peaks that agree with the SPOC-generated
light curve.

Further analysis using {\it TESS}-Localize provides some support to our conclusion that the lower frequencies arise from this star. The fits are not significant, but have $p < 0.2$, suggesting that this conclusion may be somewhat likely.

\section{Discussion of individual stars} \label{ap_B}

Here we discuss each star according to their positions in the final H-R diagram, starting from the hottest objects and progressing towards lower \teff. We give arguments about whether the given objects can be considered members of an established class of variable stars, or whether they cannot be explained this way.\\

\subsection{Stars in the $\beta$ Cephei domain}

TIC 202567458 = HD 332044: most of the Fourier spectrum of this star's pulsations can be described by a basic multitude of signals between 2.91--3.24\,d$^{-1}$ and a single frequency of 1.46\,d$^{-1}$ as well as a surprisingly long series of peaks at combination multiples thereof. Pressure modes are hard to discern. In addition, there is a 1/f "noise floor" reminiscing the signals usually connected with Internal Gravity Waves (IGWs, e.g., see \citealt{2023A&A...674A.134R}). Given that the star lies close to the zero-age main sequence in Fig.\,\ref{final_position}, we conclude that the apparently coherent signals are due to gravity mode pulsations and their combination frequencies \citep{2015MNRAS.450.3015K}. Thus there is no reason for calling this object a "Maia" variable.

TIC 312637783 = HD 228456: this star shows multiple pulsation frequencies between 8--9\,d$^{-1}$ and 11\,d$^{-1}$ as well as some combination frequencies thereof. It has a high RUWE parameter of 3.53 (cf. Table \ref{table2_atmpars}). With this and the position of the star in Fig.\,\ref{final_position}, we classify this star as a suspected binary containing at least one $\beta$ Cephei pulsator. There is nothing unusual about this object.

TIC 374038418 = BD+60 416: the Fourier spectrum is dominated by three coherent frequencies typical of $\beta$ Cephei pulsators. Superposed on that is some non-coherent variability which could be due to IGWs or mass loss. All these phenomena are consistent with this object's temperature and luminosity put it in the $\beta$ Cephei instability strip near the end of the main sequence phase. There is no need to invoke a possible "Maia" phenomenon.

TIC 377099704 = HD 177195: this star is located at the red edge of the $\beta$ Cephei instability strip and inside the instability strip of the Slowly Pulsating B (SPB stars). Its pulsation spectrum is very complicated, with many modes in both the expected p and g mode frequency domains that do not seem to be harmonically related. We classify it as a "hybrid" $\beta$ Cep/SPB pulsator.

TIC 251196433 = BD+57 655: this is a very rapidly rotating star with a large RUWE value suggestive of binarity. In Sect.\ref{sec:rv} we also commented on its line-profile shape atypical of a single star. This object is located near the edges of both the $\beta$ Cep and SPB instability strips. Its pulsation spectrum is dominated by an apparent p mode, and has many frequencies in both the $\beta$ Cep and SPB domains plus some combination frequencies. According to us this is another "hybrid" $\beta$ Cep/SPB pulsator. 

\subsection{Stars in the SPB domain}

TIC 354793407 = HD 21699: this star's light curve is dominated by a 2.5-d variation that is linked to rotation \citep{1987AJ.....94..737S}. Superposed on that is low-amplitude multiperiodic variability with periods around 40 minutes, typical for unevolved $\delta$~Scuti stars. At $T_{\rm eff}=16000$\,K, this star is too hot to be a $\delta$~Scuti star (cf. Table \ref{table2_atmpars}). However, it also has a large RUWE of 1.81, suggestive of binarity. A hypothesized main sequence $\delta$~Scuti companion would be expected to have about 1/60 of the luminosity of the chemically peculiar primary. Therefore the small ($<0.23$\,mmag) short-period pulsation amplitudes in the {\it {\it TESS}} light curve would have intrinsic values of about 14 mmag and less, quite plausible for such an object. We therefore suspect that a binary companion is responsible for the short-period pulsations.

TIC 36557487 = HD 49643: similar to TIC 202567458, this star shows groups of closely spaced frequencies between 1.98 and 3.13\,d$^{-1}$ and harmonics thereof, that can be discerned out to 37\,d$^{-1}$ in the Sector 6 data and that we analogically classify as g mode pulsations. Apart from that, in {\it {\it TESS}} Sector 33 the light curve shows a sudden spectacular increase in total pulsation amplitude by a factor of about 15 in connection with a brightening event. We, therefore, suspect this is a hitherto unrecognized Be star as discussed e.g. by \citep{2022AJ....163..226L}. In any case, we find no reason why this object should be a "Maia" variable.

TIC 424032547 = HD 350990: the light curve of this object is dominated by a frequency of 5.025\,d$^{-1}$ whose amplitude is different within the three sectors it was observed. Interestingly, the amplitude of this variation is {\it inversely} correlated with the CROWDSAP parameter, i.e. the larger the flux contribution of the target in the photometric aperture, the lower the amplitude. This suggests that this object is not the source of the variability (cf. Sect.\ref{tic4240}). Furthermore, careful prewhitening of the dominant frequency and its harmonics reveals the presence of a weak subharmonic. A phase fold of the light curve around the subharmonic frequency yields a light curve shape indicative of an ellipsoidal variable.

TIC 236785664 = HD 175511 is classified as a late-type Be star. The Fourier spectrum of its {\it {\it TESS}} light curve shows a single frequency near 1.24\,d$^{-1}$ and a multitude of peaks between 12 and 43\,d$^{-1}$ that are not harmonically related. Such pulsation frequencies would be typical for a $\delta$ Scuti star. At $T_{\rm eff}$ = 12500\,K  this star is much hotter than such pulsators. Even its rapid rotation $v \sin i = 260$\,\kms should not create sufficient oblateness that its equatorial regions would be cool enough to excite pulsations. Only its Gaia RUWE parameter (1.20) gives a hint of possible binarity, which by itself we do not find sufficient to rule this object out as a "Maia" candidate. One may suspect that this star could be a case related to HD 42477. However, an examination for possible coupling between the g and p modes yielded a null result.

TIC 192780911 = HD 19921: the {\it {\it TESS}} data show little intrinsic variability except for a weak signal at 3.9\,d$^{-1}$, which could be an SPB-type pulsation modified by the rapid rotation of the star ($v \sin i = 273$\,\kms). The RUWE value of 1.77 is suggestive of binarity. 

TIC 16781787 =  HD 207674: this object's variability is predominantly singly periodic with a frequency of 6.0\,d$^{-1}$. Two very weak signals near 2.35\,d$^{-1}$ are present in the S56 data, but are not subharmonics. As the strongest frequency is too high to be caused by rotation or binarity we speculate that this star's very rapid rotation shifts the dominant pulsation frequency with an estimated amount of $>2$\,d$^{-1}$ to such a large value for an SPB star.

\subsection{Stars between the SPB and $\delta$ Scuti strips}

TIC 64774351 = HD 204905: the frequency spectrum of this star changed between the Sector 15/16 and 56/57 observations, but the range of dominating frequencies remained around 4 -- 5\,d$^{-1}$. The Fourier Transform contains peaks between 1 -- 31\,d$^{-1}$ that do not appear to cluster in groups. However, a closer look reveals that the higher frequencies can in all cases be explained by combinations of lower frequencies, with these few base frequencies located in the g mode domain. We therefore surmise that this star is related to those with frequency groupings. The high-frequency signals are then not intrinsic normal modes of pulsation but combination frequencies. 

TIC 26369887 =  HD 222142: there is a frequency group between 3.05--3.5\,d$^{-1}$ that, together with a single frequency of 2.47\,d$^{-1}$, appears to form all other observed frequencies by linear combinations. This object then is a very cool SPB star. Its rapid rotation could have an influence on pulsational excitation. There are no high frequencies that would make this object a "Maia" candidate.

TIC 83803744 = HD 1009 is a HgMn star that has two periods consistently present in the {\it {\it TESS}} data sets, one at 0.89\,d$^{-1}$ that is apparently the first harmonic of its rotation period \citep{2021MNRAS.506.5328K} and a low amplitude (0.04 - 0.05 mmag) apparent pulsation at 12.36\,d$^{-1}$. Even though our two radial velocity measurements are consistent within the errors, it is known that HgMn stars often occur in binary systems (e.g., \citealt{1985A&A...146..341G}) which is why we are reluctant to classify the apparent pulsation as "Maia" variability: the pulsations could arise from a $\delta$ Scuti companion.

TIC 20534584 = HD 176052: the low-frequency part of the Fourier Transform of the {\it {\it TESS}} light curve of this fairly rapid rotator ($v \sin i = 150$\,\kms) is constituted by harmonically related frequencies based on a group around 3.35--3.40\,d$^{-1}$. A single significant higher frequency peak occurs at 22.64\,d$^{-1}$ with an amplitude of only 0.16 mmag (the peak at 27.69\,d$^{-1}$ is not significant). Given the high RUWE value of this object (2.26), we suspect that the high-frequency variation comes from a companion of the $\delta$ Scuti type.

TIC 352781227 = HD 351435: this is a very rapid rotator with a consequently large uncertainty in its effective temperature (cf. Table \ref{table2_atmpars}). The pulsation spectra are clearly variable in amplitude from sector to sector of data and even within single sectors, but the underlying frequencies are essentially the same, suggesting these are unresolved groups of frequencies. The signals below 10\,d$^{-1}$ can be fairly easily explained with combinations of two basic frequencies near 1.43 and 3.84\,d$^{-1}$, and the higher amplitudes of the combination frequencies with heavy geometrical cancellation of the base frequencies \citep[see][]{2015MNRAS.450.3015K}. The single higher frequency signal at 13.66\,d$^{-1}$ can be matched with a third-order combination frequency and may also be an unresolved group. Thus this object could be a pure g mode pulsator in spite of the observed high frequencies or a p mode that is driven in the equatorial regions of this oblate star\citep{2023MNRAS.521.4765K}.

TIC 434340780 = BD+61 2380 is a rapidly rotating star of late B spectral type. Its two pulsation frequencies are too low for p modes at this star's position in the HR Diagram, which is why we interpret them in terms of g modes shifted to high frequency by rapid rotation ($f_{\rm rot} \approx 2$\,d$^{-1}$).

\subsection{Stars in the $\delta$ Scuti strip}

The behaviour of all these objects is consistent with that of $\delta$ Scuti pulsators, which is also consistent with their positions in the HR Diagram. Nevertheless, they will be discussed briefly in what follows.

TIC 201739321 = HD 356 is located at the blue edge of the $\delta$ Scuti instability strip. The short-period variations can therefore be classified as $\delta$ Scuti pulsations. The two signals below 10\,d$^{-1}$ are harmonically related.

TIC 317512446 = HD 220855 is located just inside the $\delta$ Scuti instability strip. This is consistent with its two short-period pulsation frequencies.

TIC 27804376 = HD 210661 and TIC 26368855 = HD 222017 are two multiperiodic $\delta$ Scuti stars near the hot border of their pulsational instability strip with intermediate projected rotational velocities.

TIC 255996252 = HD 196358 has similar characteristics to the two stars just discussed, but its pulsation spectrum is not as rich.

TIC 450085823 = HD 287149 shows multiple pulsation frequencies of the $\delta$ Scuti type below 20\,d$^{-1}$. Its pulsation spectrum exhibits considerable amplitude changes between its two {\it {\it TESS}} light curves taken two years apart.

TIC 402910080 = HD 193989 is a very rapidly rotating star with a wide range of $\delta$ Scuti type pulsation frequencies observed. Amplitude changes between the different epochs of {\it {\it TESS}} observations are easily discernable.

TIC 387757610 = HD 200506 is another rapidly rotating $\delta$ Scuti star. Its pulsation spectrum appears fairly stable over time and it also exhibits g mode pulsations of the $\gamma$ Doradus type plus combination frequencies thereof.

TIC 377443211 = HD 185757 also rotates rapidly. It shows a wide range of $\delta$ Scuti type pulsation frequencies whose amplitudes somewhat change over time.

TIC 90647390 = HD 278751 shows a few $\delta$ Scuti type pulsations and some combination frequencies thereof. Its Gaia RUWE value strongly suggests binarity.

TIC 467219052 = HD 239419 is the slowest pulsating $\delta$ Scuti star in our sample, and it is also farthest away from the zero-age main sequence. Its pulsation amplitudes vary somewhat over time.

TIC 367351021 = HD 216486 has the highest RUWE value among our targets and a set of almost equally spaced $\delta$ Scuti type pulsation frequencies.

TIC 50850670 = BD+65 291 belongs to the $\delta$ Scuti stars of intermediate rotational velocities. It has three distinct groups of pulsation frequencies, the lowest frequency group likely due to $\gamma$ Doradus type pulsations. The two other groups are separated by a factor of two in frequency but the individual modes are not harmonically related.

TIC 335484421 = BD+62 2044 is the coolest $\delta$ Scuti star in our sample, with a wide range and a large number of pulsation frequencies.

\setcounter{table}{0}
\renewcommand{\thetable}{A\arabic{table}}

\begin{table*}
    \centering
\caption{Radial velocity measurements. * represents the cannot measure radial velocity value due to low signal-to-noise level.}\label{rv_table} 
\begin{tabular}{c|cl|cl}
    \hline
    &\multicolumn{2}{c|}{\hrulefill FIES\,\hrulefill}  &\multicolumn{2}{c}{\hrulefill \,HERMES\,\hrulefill}\\
    TIC      & JD                  & v$_{r}$        &  JD                  & v$_{r}$      \\
             &  2458000+                   & (km/s)        & 2459000+                 &  (km/s)    \\

\hline
    16781787 &  1122.4625   & 7.33\,$\pm$\,2.52  & 1057.6369  & 3.44\,$\pm$\,1.46  \\ 
    20534584 &              &                    &  996.4778  & -13.68\,$\pm$\,2.22 \\ 
             &              &                    & 993.5008   & -14.38\,$\pm$\,2.15  \\
             &              &                    & 996.4778   & -13.2\,$\pm$\,2.53   \\
    26368855 & 1112.5543    & -13.98\,$\pm$\,2.56& 1054.7315  & -6.06\,$\pm$\,1.35   \\
             & 1112.4515    & -6.81\,$\pm$\,2.80 & 1056.7014  & -6.40\,$\pm$\,1.32   \\
    26369887 & 1112.5438    & -2.02\,$\pm$\,0.65 & 1057.7031  & -1.36\,$\pm$\,0.85  \\
    27804376 & 1120.3924    &-9.30\,$\pm$\,1.69  & 1052.6768  & -17.65\,$\pm$\,1.85   \\
             & 1137.5354    & -11.50\,$\pm$\,2.16&            &                    \\
    36557487 &              &                    & 1190.6814  & 17.95\,$\pm$\,2.53  \\
             &              &                    & 1190.6814  & 19.86\,$\pm$\,2.86  \\
    50850670 & 1123.6194    & -12.80\,$\pm$\,1.86& 1230.4578  & 5.41\,$\pm$\,1.85   \\
             & 1152.5264    &-9.50\,$\pm$\,1.33  & 1237.4024  & 4.67\,$\pm$\,2.74    \\
    64774351 & 1121.4633    & -9.03\,$\pm$\,2.18 & 780.5811   & -9.95\,$\pm$\,2.37  \\
             &              & -4.21\,$\pm$\,1.20 & 1083.6384  & -9.50\,$\pm$\,3.85  \\
             &              & -5.48\,$\pm$\,1.28 &            &       \\
    83803744 & 1112.5647    & -14.19\,$\pm$\,2.93&    &   \\ 
             & 1113.6432    & -14.00\,$\pm$\,3.71&    &   \\ 
    90647390 & 1123.6362    & -50.00\,$\pm$\,3.41& 1245.4465   & -32.57\,$\pm$\,4.27  \\ 
     & 1152.5466    & -38.72\,$\pm$\,3.52&  *          &   \\ 
    192780911& 1130.6064    &-14.39\,$\pm$\,2.45 & 1164.5964   & 0.15\,$\pm$\,2.80   \\
             &              &                    & 1194.5904   & 2.44\,$\pm$\,2.04   \\
    201739321& 1112.5710    & -20.78\,$\pm$\,3.11& 1056.7346   & -9.75\,$\pm$\,3.21  \\
             & 1112.4675    & -21.48\,$\pm$\,4.83&    &\\
             & 1113.6515    & -20.78\,$\pm$\,3.11&    &  \\
    202567458 & &   &   &   \\
    236785664 &             &                     &625.4982  & -15.50\,$\pm$\,2.56   \\
              & &                                 &660.5974  &-4.41\,$\pm$\,1.22  \\
              & &                                 &990.4414  &-7.01\,$\pm$\,1.32   \\
              & &                                 &991.4537  &-6.98\,$\pm$\,1.11   \\
    251196433 & 1153.4524 &-30.18\,$\pm$\,3.18    &1147.6737 &-33.63\,$\pm$\,3.89   \\
              &           &                       &1194.4703 &-30.51\,$\pm$\,3.69   \\
    255996252 & 1114.4288 &7.47\,$\pm$\,2.15      &997.7101  &9.39\,$\pm$\,1.26   \\
              & 1120.3858 &8.04\,$\pm$\,1.28      &997.7101  &10.70\,$\pm$\,1.36   \\
              & 1122.4441 &7.04\,$\pm$\,1.84      &          &   \\
    312637783 & 1121.4469 &11.62\,$\pm$\,2.10     &   &   \\ 
              & 1122.4149 &8.42\,$\pm$\,1.83     &   &   \\ 
              & 1121.7385 &5.68\,$\pm$\,1.38     &   &   \\ 
    317512446 & 1107.5420 & 9.40\,$\pm$\,2.22    &   &   \\
              & 1107.6184 & 9.86\,$\pm$\,1.78  &   &   \\
              & 1083.6451 & 3.04\,$\pm$\,3.52  &   &   \\

      \hline
\end{tabular}
\end{table*}

\setcounter{table}{0}
\begin{table*}
    \centering
\caption{Continuation.} \label{rv_table}
\begin{tabular}{c|cl|cl}
    \hline
    &\multicolumn{2}{c|}{\hrulefill FIES\,\hrulefill}  &\multicolumn{2}{c}{\hrulefill \,HERMES\,\hrulefill}\\
    TIC      & JD                  & v$_{r}$        &  JD                  & v$_{r}$      \\
             &  2458000+                   & (km/s)        & 2459000+                 &  (km/s)    \\

\hline
    335484421 & 1113.4823 & 0.26\,$\pm$\,1.02  &   &   \\
              & 1123.3351 & 0.77\,$\pm$\,0.98  &   &   \\
  352781227 & &         &1056.5846   & 25.77\,$\pm$\,5.32   \\
              & &         &1056.6060   & 31.85\,$\pm$\,5.58   \\
    354793407 & 1131.4274 &-26.78\,$\pm$\,2.15   &1191.4345   & 4.77\,$\pm$\,1.58   \\
              & 1121.5663 &-24.13\,$\pm$\,2.27   &1194.5965   & 5.29\,$\pm$\,1.62   \\
    367351021 & 1120.3990 &-60.00\,$\pm$\,5.75   &1260.4277   &-63.91\,$\pm$\,5.79   \\
              & 1152.5566 &-63.71\,$\pm$\,5.79   &            &   \\
    374038418 &  1120.3990 &-60.00\,$\pm$\,5.75   &1260.4277   &-63.91\,$\pm$\,5.79   \\
              & 1152.5566 &-63.71\,$\pm$\,5.79   &            &   \\
              & 1242.3445 &-51.71\,$\pm$\,5.14   &            &   \\
    377099704 & &   & 652.4951  &-28.85\,$\pm$\,2.18   \\
              & &   & 654.4107  &-28.34\,$\pm$\,2.08   \\
              & &   & 668.4067  &-29.56\,$\pm$\,1.85   \\
              & &   & 990.4497  &-30.78\,$\pm$\,1.98   \\
              & &   & 991.4639  &-27.11\,$\pm$\,2.31   \\
    377443211 & 1114.4406& -6.31\,$\pm$\,1.89  &997.7053   & -18.21\,$\pm$\,4.89  \\
              & 1122.3990& -8.88\,$\pm$\,2.50  &           &   \\
    387757610 & 1119.3948& -16.81\,$\pm$\,2.80  &997.7178  &-14.00\,$\pm$\,2.41   \\
              & 1122.4504& -14.48\,$\pm$\,2.86  &1083.6543 &-10.24\,$\pm$\,5.12   \\
    402910080 & 1119.3948 & 22.98\,$\pm$\,4.17   &   &   \\
              & 1122.4344 & 16.76\,$\pm$\,5.41   &   &   \\
    424032547 &1122.4045 & 13.06\,$\pm$\,3.75   &   &   \\
              & 1138.4108 & 14.38\,$\pm$\,4.20   &   &   \\
    434340780 & 1120.3760 & -7.39\,$\pm$\,4.81   &   &   \\
              & 1153.3448 & -3.75\,$\pm$\,5.13   &   &   \\
    450085823 & 1112.7434 & 10.25\,$\pm$\,5.75   &   &   \\
    467219052 & 1113.4962 & 1.68\,$\pm$\,1.12   & 1056.6205   & 4.17\,$\pm$\,2.15   \\
    \hline
\end{tabular}
\end{table*}

\setcounter{table}{1}
\begin{table*}
 \begin{scriptsize}
\caption{The result of the analysis of chemical abundances. The numbers given in the brackets represent the number of used lines in the analysis.}\label{table_abundance}
\begin{tabular}{l|l|l|l|l|l|l} 
\hline\noalign{\smallskip}
&\multicolumn{1}{c}{\hrulefill TIC\,20534584\hrulefill} & \multicolumn{1}{c}{\hrulefill TIC\,26368855\hrulefill} & \multicolumn{1}{c}{\hrulefill TIC\,27804376\hrulefill}&\multicolumn{1}{c}{\hrulefill TIC\,50850670\hrulefill} &\multicolumn{1}{c}{\hrulefill TIC\,83803744\hrulefill} & \multicolumn{1}{c}{\hrulefill TIC\,90647390\hrulefill} \\  
   \hline\noalign{\smallskip}
Element   & Abundance             &  Abundance             & Abundance              & Abundance              & Abundance              & Abundance               \\
\hline
$_{6}$C   &                       & 8.15\,$\pm$\,0.33 (2)  & 8.41\,$\pm$\,0.23 (6)  &                         &                        & 8.58\,$\pm$\,0.29 (4) \\
$_{7}$N   &                       &                        &                        &                         &                        &                       \\
$_{8}$O   &                       & 8.98\,$\pm$\,0.33 (2)  &                        &                         &                        &                             \\
$_{11}$Na &                       &                        & 6.45\,$\pm$\,0.42 (1)  & 6.47\,$\pm$\,0.26 (2)   &                        &                          \\
$_{12}$Mg &7.75\,$\pm$\,0.46 (2)  & 7.88\,$\pm$\,0.33 (2)  & 7.94\,$\pm$\,0.52 (5)  & 8.02\,$\pm$\,0.39 (4)   & 6.53\,$\pm$\,0.31 (1)  & 7.93\,$\pm$\,0.27 (4)  \\
$_{14}$Si &7.26\,$\pm$\,0.46 (2)  & 7.62\,$\pm$\,0.37 (3)  & 7.30\,$\pm$\,0.53 (4)  & 7.02\,$\pm$\,0.36 (18)  & 7.85\,$\pm$\,0.29 (2)  & 7.17\,$\pm$\,0.37 (3)  \\
$_{16}$S &                       &                        &                        & 7.39\,$\pm$\,0.28 (2)   &                        & 7.06\,$\pm$\,0.34 (1)   \\
$_{20}$Ca &6.86\,$\pm$\,0.45 (2)  & 6.43\,$\pm$\,0.43 (13) & 6.79\,$\pm$\,0.47 (8)  & 5.86\,$\pm$\,0.27 (15)  & 6.53\,$\pm$\,0.30 (2)  & 6.71\,$\pm$\,0.23 (8)  \\
$_{21}$Sc & 3.69\,$\pm$\,0.45 (2) & 3.19\,$\pm$\,0.36 (3)  & 3.22\,$\pm$\,0.35 (3)  & 2.58\,$\pm$\,0.33 (5)   & 3.28\,$\pm$\,0.33 (2)  & 3.27\,$\pm$\,0.26 (2)  \\
$_{22}$Ti &4.17\,$\pm$\,0.36 (7)  & 4.97\,$\pm$\,0.40 (17) & 5.19\,$\pm$\,0.38 (12) & 4.72\,$\pm$\,0.25 (34)  & 4.53\,$\pm$\,0.26 (5)  & 5.31\,$\pm$\,0.24 (18) \\
$_{23}$V  &                       &                        & 4.44\,$\pm$\,0.22 (2)  & 4.85\,$\pm$\,0.33 (6)   & 5.00\,$\pm$\,0.34 (2)  &                       \\
$_{24}$Cr &5.10\,$\pm$\,0.36 (6)  & 5.80\,$\pm$\,0.31 (10) & 5.84\,$\pm$\,0.25 (12) & 5.69\,$\pm$\,0.25 (33)  & 6.19\,$\pm$\,0.27 (11) & 6.01\,$\pm$\,0.29 (10) \\
$_{25}$Mn &                       &                        & 		            & 5.45\,$\pm$\,0.33 (10)  & 7.32\,$\pm$\,0.35 (2)  & 6.31\,$\pm$\,0.35 (2) \\
$_{26}$Fe &6.83\,$\pm$\,0.34 (15) & 7.44\,$\pm$\,0.25 (33) & 7.58\,$\pm$\,0.35 (42) & 7.64\,$\pm$\,0.18 (104) & 7.11\,$\pm$\,0.37 (14) & 7.53\,$\pm$\,0.27 (62)\\
$_{27}$Co &                       &                        &                        & 5.72\,$\pm$\,0.26 (2)   &                        &                             \\
$_{28}$Ni &                       & 6.27\,$\pm$\,0.34 (5)  & 6.24\,$\pm$\,0.44 (11) & 6.61\,$\pm$\,0.21 (33)  & 5.86\,$\pm$\,0.36 (1)  & 6.30\,$\pm$\,0.30 (4)  \\
$_{29}$Cu &                       &                        &                        & 4.38\,$\pm$\,0.27 (1)   &                        &                          \\
$_{30}$Zn &                       &                        &                        & 4.43\,$\pm$\,0.26 (1)   &                        &                         \\
$_{38}$Sr &                       & 2.76\,$\pm$\,0.33 (1)  &                        & 3.39\,$\pm$\,0.26 (2)   & 2.44\,$\pm$\,0.32 (1)  &                        \\
$_{39}$Y  &3.06\,$\pm$\,0.46 (1)  & 2.45\,$\pm$\,0.33 (2)  & 2.43\,$\pm$\,0.42 (2)  & 2.67\,$\pm$\,0.31 (6)   & 3.39\,$\pm$\,0.36 (1)  & 2.24\,$\pm$\,0.31 (2)  \\
$_{40}$Zr &                       &                        & 2.89\,$\pm$\,0.42 (1)  & 2.96\,$\pm$\,0.26 (4)   &                        &                        \\
$_{56}$Ba &                       & 2.14\,$\pm$\,0.33 (2)  & 3.10\,$\pm$\,0.42 (1)  &                         &                        & 2.61\,$\pm$\,0.33 (2) \\
\hline\noalign{\smallskip}
\end{tabular}
 \end{scriptsize}
\end{table*}

 \newpage

\setcounter{table}{1}

\begin{table*}
 \begin{scriptsize}
\caption{Continuation.}
\begin{tabular}{l|l|l|l|l|l|l} 
\hline\noalign{\smallskip}
&\multicolumn{1}{c}{\hrulefill TIC\,201739321\hrulefill} & \multicolumn{1}{c}{\hrulefill TIC\,255996252\hrulefill} & \multicolumn{1}{c}{\hrulefill TIC\,312637783\hrulefill}&\multicolumn{1}{c}{\hrulefill TIC\,317512446\hrulefill} &\multicolumn{1}{c}{\hrulefill TIC\,335484421\hrulefill} & \multicolumn{1}{c}{\hrulefill TIC\,354793407\hrulefill} \\  
   \hline\noalign{\smallskip}   
$_{6}$C   & 			            &  8.19\,$\pm$\,0.35 (13) & 8.39\,$\pm$\,0.32 (29)  & 8.94\,$\pm$\,0.27 (7)  & 8.78\,$\pm$\,0.41 (6)  &	\\ 		   
$_{7}$N   & 			            &  7.45\,$\pm$\,0.24 (2)  & 7.87\,$\pm$\,0.36 (21)  & 8.03\,$\pm$\,0.28 (1)  &			              &	\\ 		   
$_{8}$O   & 			            &  8.79\,$\pm$\,0.24 (2)  &		      	            &			             &			              &	\\ 		   
$_{11}$Na & 			            &  6.50\,$\pm$\,0.26 (2)  &		      	            &			             & 6.44\,$\pm$\,0.46 (2)  &	\\ 		   
$_{12}$Mg &  7.72\,$\pm$\,0.24 (5)  &  7.65\,$\pm$\,0.34 (7)  & 8.58\,$\pm$\,0.46 (1)	& 8.10\,$\pm$\,0.30 (8)  & 7.71\,$\pm$\,0.39 (4)  & 7.64\,$\pm$\,0.35 (3) \\ 
$_{14}$Si &  7.29\,$\pm$\,0.31 (2)  &  7.49\,$\pm$\,0.23 (10) & 7.64\,$\pm$\,0.44 (7)	& 7.33\,$\pm$\,0.26 (6)  & 7.40\,$\pm$\,0.26 (12) & 8.19\,$\pm$\,0.25 (15)\\ 
$_{16}$S & 			                &  7.65\,$\pm$\,0.24 (2)  & 7.01\,$\pm$\,0.43 (8) 	& 7.50\,$\pm$\,0.36 (2)  & 6.72\,$\pm$\,0.34 (9)  &  \\ 
$_{20}$Ca &  6.22\,$\pm$\,0.30 (7)  &  6.75\,$\pm$\,0.32 (10) &		      		        & 6.81\,$\pm$\,0.27 (13) & 6.66\,$\pm$\,0.32 (23) & 6.15\,$\pm$\,0.41 (3) \\ 
$_{21}$Sc &  2.84\,$\pm$\,0.32 (4)  &  3.57\,$\pm$\,0.32 (7)  &		      		        & 3.59\,$\pm$\,0.34(3)   & 3.69\,$\pm$\,0.36 (4)  &			   \\ 
$_{22}$Ti &  4.73\,$\pm$\,0.29 (12) &  5.17\,$\pm$\,0.39 (25) &		      		        & 5.47\,$\pm$\,0.35 (36) & 5.08\,$\pm$\,0.34 (35) & 6.47\,$\pm$\,0.39 (4) \\ 
$_{23}$V  & 			            &  4.36\,$\pm$\,0.38 (5)  &		      	            & 4.43\,$\pm$\,0.39 (2)  & 5.09\,$\pm$\,0.38 (4)  & 5.29\,$\pm$\,0.39 (1) \\ 
$_{24}$Cr &  5.53\,$\pm$\,0.26 (10) &  5.82\,$\pm$\,0.33 (22) &		      		        & 5.97\,$\pm$\,0.29 (22) & 5.93\,$\pm$\,0.23 (35) & 6.75\,$\pm$\,0.30 (13)\\ 
$_{25}$Mn &  5.05\,$\pm$\,0.31 (1)  &  5.51\,$\pm$\,0.24 (2)  &		      		        & 5.53\,$\pm$\,0.37 (2)  & 5.68\,$\pm$\,0.36 (11) & 6.55\,$\pm$\,0.32 (2) \\ 
$_{26}$Fe &  7.26\,$\pm$\,0.25 (23) &  7.61\,$\pm$\,0.30 (75) & 7.46\,$\pm$\,0.43 (26)	& 7.79\,$\pm$\,0.22 (98) & 7.77\,$\pm$\,0.26 (116)& 8.19\,$\pm$\,0.26 (88)\\ 
$_{27}$Co & 			            &  		                  &		      	            &			             & 5.98\,$\pm$\,0.48 (1)  &			   \\ 
$_{28}$Ni &  6.14\,$\pm$\,0.33 (2)  & 6.38\,$\pm$\,0.39 (19)  & 7.23\,$\pm$\,0.46 (2)	& 6.60\,$\pm$\,0.32 (6)  & 6.78\,$\pm$\,0.30 (31) & 7.00\,$\pm$\,0.35 (8) \\ 
$_{29}$Cu & 			            &  		                  &		      	            &			             & 3.93\,$\pm$\,0.46 (1)  &			   \\ 
$_{30}$Zn & 			            &  		                  &		      	            &			             & 4.72\,$\pm$\,0.46 (1)  &			   \\ 
$_{38}$Sr & 			            & 3.71\,$\pm$\,0.24 (1)   &		      	            & 4.17\,$\pm$\,0.33 (1)  & 4.22\,$\pm$\,0.36 (2)  &			   \\ 
$_{39}$Y  & 			            & 2.79\,$\pm$\,0.31 (5)   &		      	            & 2.80\,$\pm$\,0.38 (1)  & 3.01\,$\pm$\,0.35 (3)  &			   \\ 
$_{40}$Zr &                         & 3.14\,$\pm$\,0.36 (1)   & 3.34\,$\pm$\,0.29 (3)   &		      	         &  3.92\,$\pm$\,0.37 (1) & 3.48\,$\pm$\,0.39 (3) 			   \\ 
$_{56}$Ba &                         &			              & 3.18\,$\pm$\,0.24 (2)   &		      	         &   2.64\,$\pm$\,0.38 (2)& 3.74\,$\pm$\,0.42 (2) 			   \\ 
\hline\noalign{\smallskip}
\end{tabular}
 \end{scriptsize}
\end{table*}

\setcounter{table}{1}
\begin{table*}
 \begin{scriptsize}
\caption{Continuation.}
\begin{tabular}{l|l|l|l|l|l|l} 
\hline\noalign{\smallskip}
&\multicolumn{1}{c}{\hrulefill TIC\,367351021\hrulefill} & \multicolumn{1}{c}{\hrulefill TIC\,374038418\hrulefill} & \multicolumn{1}{c}{\hrulefill TIC\,377099704\hrulefill}&\multicolumn{1}{c}{\hrulefill TIC\,377443211\hrulefill} &\multicolumn{1}{c}{\hrulefill TIC\,387757610\hrulefill} & \multicolumn{1}{c}{\hrulefill TIC\,467219052\hrulefill} \\  
   \hline\noalign{\smallskip}  
Element    & Abundance  	    & Abundance 	     & Abundance	      & Abundance	       & Abundance		 & Abundance			   \\
\hline
$_{6}$C    & 8.38\,$\pm$\,0.136 (13)&			              & 8.31\,$\pm$\,0.28 (12) & 		                &  8.30\,$\pm$\,0.37 (6) & 8.31\,$\pm$\,0.46 (5)\\
$_{7}$N    & 8.15\,$\pm$\,0.26 (1)  &			              & 8.09\,$\pm$\,0.30 (12) & 		                & 		                 &		       \\
$_{8}$O    & 8.96\,$\pm$\,0.26 (2)  &			              &  		               & 8.90\,$\pm$\,0.35 (2)  &  8.73\,$\pm$\,0.28 (4) &  8.83\,$\pm$\,0.30 (1)\\
$_{11}$Na  & 6.72\,$\pm$\,0.26 (2)  &			              &  		               & 		                &  6.83\,$\pm$\,0.37 (2) &  6.36\,$\pm$\,0.30 (1)\\
$_{12}$Mg  & 8.03\,$\pm$\,0.33 (5)  & 7.43\,$\pm$\,0.56 (1)  & 8.32\,$\pm$\,0.37 (5)   & 7.90\,$\pm$\,0.36 (5)  &  7.99\,$\pm$\,0.33 (5) &  7.73\,$\pm$\,0.38 (3)\\
$_{14}$Si  & 7.64\,$\pm$\,0.43 (17) &			             & 7.01\,$\pm$\,0.31 (18)  & 8.05\,$\pm$\,0.31 (3)  &  7.46\,$\pm$\,0.40 (6) &  7.36\,$\pm$\,0.30 (5)\\
$_{16}$S   & 7.44\,$\pm$\,0.26 (2)                       &   &			               &			            & 8.26\,$\pm$\,0.34 (2) & 7.06\,$\pm$\,0.30 (1)\\
$_{20}$Ca  & 6.58\,$\pm$\,0.32 (15) & 6.96\,$\pm$\,0.57 (1)  &  		               & 6.31\,$\pm$\,0.31 (5)  &  6.42\,$\pm$\,0.33 (7) & 6.36\,$\pm$\,0.38 (12)\\
$_{21}$Sc  & 3.23\,$\pm$\,0.28 (7)  &			             &  		               & 2.91\,$\pm$\,0.34 (2)  &  2.99\,$\pm$\,0.30 (4) &  3.22\,$\pm$\,0.26 (5)\\
$_{22}$Ti  & 5.12\,$\pm$\,0.32 (23) &			             &  		               & 5.02\,$\pm$\,0.37 (6)  &  5.15\,$\pm$\,0.36 (10)&  5.23\,$\pm$\,0.39 (12)\\
$_{23}$V   &			            &			             &			               &  		                & 		                 &  4.19\,$\pm$\,0.30 (1)\\
$_{24}$Cr  & 5.80\,$\pm$\,0.40 (21) &			             &  		               & 5.66\,$\pm$\,0.32 (6)  &  5.60\,$\pm$\,0.34 (9) &  5.82\,$\pm$\,0.34 (10) \\
$_{25}$Mn  & 5.55\,$\pm$\,0.25 (3)  &			             &  		               & 		                &  5.35\,$\pm$\,0.34 (2) &  \\
$_{26}$Fe  & 7.54\,$\pm$\,0.30 (86) & 8.53\,$\pm$\,0.46 (2)  & 7.31\,$\pm$\,0.31 (19) & 7.31\,$\pm$\,0.30 (30)  &  7.40\,$\pm$\,0.36 (18)&  7.49\,$\pm$\,0.30 (40)\\
$_{27}$Co  &			            &			             &			              &  			            &		                 &  \\
$_{28}$Ni  & 6.34\,$\pm$\,0.39 (29) & 7.47\,$\pm$\,0.48 (1)  &  		              & 		                & 6.28\,$\pm$\,0.33 (5)  &  6.20\,$\pm$\,0.30 (8)\\
$_{29}$Cu  &			            &			             &			              &  		                & 		                 & \\
$_{30}$Zn  & 5.92\,$\pm$\,0.34 (1)  &			             &  		              & 		                & 		                 &  4.99\,$\pm$\,0.30 (1)\\
$_{38}$Sr  &			            &			             &			              &  		                & 3.41\,$\pm$\,0.34 (1)  &  2.76\,$\pm$\,0.30 (1)\\
$_{39}$Y   & 2.51\,$\pm$\,0.34 (2)  &			            &  		                  & 		                & 3.13\,$\pm$\,0.33 (2)  &  2.41\,$\pm$\,0.30 (1)\\
$_{40}$Zr  &			            &			            &			              &  		               & 		                 &  3.10\,$\pm$\,0.30 (1)\\
$_{56}$Ba  & 2.59\,$\pm$\,0.32 (3)  &			            &  		                  & 1.90\,$\pm$\,0.34 (2)  & 2.55\,$\pm$\,0.33 (2)  &  2.48\,$\pm$\,0.31 (3)\\
\hline\noalign{\smallskip}
\end{tabular}
 \end{scriptsize}
\end{table*}

\end{document}